\documentclass[11pt,paper]{JHEP3}
\usepackage{epsfig}
\usepackage{latexsym}
\usepackage{amssymb}
\usepackage{amsmath}
\usepackage{graphicx}
\usepackage{dcolumn}

\newcommand{\be}{\begin{equation}}
\newcommand{\ee}{\end{equation}}
\newcommand{\bea}{\begin{eqnarray}}
\newcommand{\eea}{\end{eqnarray}}

\newcommand{\psibar}{\overline \psi}

\newcommand{\msbar}{{\rm \overline{MS\kern-0.05em}\kern0.05em}}
\newcommand{\PM}{{\Bbb P}_M}
\newcommand{\Xop}{{\Bbb X}}
\newcommand{\R}{\rm R}
\newcommand{\wS}{\ensuremath{\tilde\rho_{\R}} }

\title{Spectral density of the Dirac operator in two-flavour QCD}

\author{Georg P.~Engel\\
Dipartimento di Fisica, Universit\`a Milano-Bicocca, \\
and INFN, Sezione di Milano-Bicocca, \\
Piazza della Scienza 3, 20126 Milano, Italy \\
E-mail: \email{georg.engel@mib.infn.it}}

\author{Leonardo Giusti\\
Dipartimento di Fisica, Universit\`a Milano-Bicocca, \\
and INFN, Sezione di Milano-Bicocca, \\
Piazza della Scienza 3, 20126 Milano, Italy \\
E-mail: \email{leonardo.giusti@mib.infn.it}}

\author{Stefano Lottini\\
DESY, Zeuthen, \\
Platanenallee 6, 15738 Zeuthen, Germany \\
E-mail: \email{stefano.lottini@desy.de}}

\author{Rainer Sommer\\
DESY, Zeuthen, \\
Platanenallee 6, 15738 Zeuthen, Germany \\
E-mail: \email{rainer.sommer@desy.de}}

\abstract{
We compute the spectral density of the (Hermitean) Dirac operator in 
Quantum Chromodynamics with two light degenerate quarks near the 
origin. We use CLS/ALPHA lattices generated with two flavours 
of $O(a)$-improved 
Wilson fermions corresponding to pseudoscalar meson masses down to $190$~MeV,
and with spacings in the range $0.05$--$0.08$~fm. Thanks to the coverage 
of parameter space, we can extrapolate our data to the 
chiral and continuum limits with confidence. The results show that the spectral 
density at the origin is non-zero because the low modes of the Dirac operator 
do condense as expected in the Banks--Casher mechanism. Within errors, the spectral density 
turns out to be a constant function up to eigenvalues 
of $\approx 80$~MeV. Its value agrees with the one extracted from the 
Gell--Mann--Oakes--Renner relation.
}

\begin{document}

\section{Introduction}
There is overwhelming evidence that
the chiral symmetry group $SU(N_f)_L\times SU(N_f)_R$ of 
Quantum Chromodynamics (QCD) with a small number $N_f$ 
of light flavours breaks spontaneously to $SU(N_f)_{L+R}$. 
This progress became possible over the last decade thanks to 
the impressive speed-up of the numerical simulations of lattice 
QCD with light dynamical fermions~\cite{Hasenbusch:2001ne,Luscher:2005rx,
DelDebbio:2006cn,Urbach:2005ji,Luscher:2012av}, for a recent compilation 
of results see~\cite{Aoki:2013ldr}. By now it is standard practice to assume this fact, 
and extrapolate phenomenologically interesting observables in the 
quark mass by applying the predictions of chiral perturbation 
theory (ChPT) \cite{Weinberg:1978kz,Gasser:1983yg}. 

The distinctive signature of spontaneous 
symmetry breaking in QCD is the set of relations among
pion masses and matrix elements which are expected to hold in 
the chiral limit~\cite{Weinberg:1978kz}. Pions 
interact only if they carry momentum, and their matrix 
elements near the chiral limit can be expressed as known 
functions of two low-energy constants (LECs), the decay constant 
$F$ and the chiral condensate $\Sigma$. 
The simplest of these relations is the Gell-Mann-Oakes-Renner (GMOR) one, 
which equals the slope of the pion mass squared with respect 
to the quark mass to $2 \Sigma/F^2$. On the one hand lattice simulations 
became so powerful that we are having now the tools to verify some of 
these relations with confidence. On the other hand very little is 
known about the dynamical mechanism which breaks
chiral symmetry. Maybe the spectrum of the Dirac operator is the 
simplest quantity to look at for an insight. Indeed many years ago Banks and Casher
suggested that chiral symmetry breaks if the low modes of the Dirac operator 
at the origin do condense and vice-versa~\cite{Banks:1979yr}. Remarkably we now 
know that 
the spectral density~\cite{Banks:1979yr,Leutwyler:1992yt,Shuryak:1992pi} 
is a renormalisable quantity to which a universal meaning can be 
assigned~\cite{Giusti:2008vb}.

The present paper is the second of two devoted to the computation of the spectral 
density of the Dirac operator in QCD with two flavours near the origin\footnote{
Preliminary results of this work were presented in 
Refs.~\cite{Engel:2013rwa,Engel:2014lat}.}. This is achieved by 
extrapolating the numerical results obtained with $O(a)$-improved Wilson fermions 
at several lattice spacings to the universal continuum limit. In the first 
paper the focus was on the physics results~\cite{Engel:2014cka}, while here we 
report the full set of results, the technical and the numerical details 
of the computation.
After fixing the notation and giving the parameters of the lattices simulated
in the second and third sections, the fourth and the fifth ones are devoted to 
two different numerical analyses of the data. Results and conclusions
are given in the last section.

\section{Spectral density of the Dirac operator}
In a space-time box of volume $V$ with periodic 
boundary conditions the spectral density of the 
Euclidean massless Dirac operator $D$ is defined as 
\be
  \rho(\lambda,m)=\frac{1}{V}\sum_{k=1}^{\infty}
  \left\langle\delta(\lambda-\lambda_k)\right\rangle\; ,
\ee
where $i\lambda_1$, $i\lambda_2$, $\ldots$ are its  
(purely imaginary) eigenvalues ordered with their 
magnitude in ascending order. As usual the bracket 
$\langle\ldots\rangle$ denotes the QCD expectation
value and $m$ the quark mass. The spectral density is a 
renormalisable observable~\cite{Giusti:2008vb,DelDebbio:2005qa}. 
Once the free parameters in the action (coupling constant and quark 
masses) have been renormalized, no renormalisation 
ambiguity is left in $\rho(\lambda,m)$. The Banks--Casher 
relation \cite{Banks:1979yr}
\be
   \lim_{\lambda \to 0}\lim_{m \to 0}\lim_{V \to \infty}\rho(\lambda,m)
   =\frac{\Sigma}{\pi}
\ee
links the spectral density to the chiral condensate
\be\label{eq:Sigrho} 
  \Sigma=-\frac{1}{2}\lim_{m \to 0}\lim_{V \to \infty}
\left\langle \bar\psi \psi\right\rangle\; ,
\ee
where $\psi$ is the quark doublet.
It can be read in either directions. If chiral symmetry is 
spontaneously broken by
a non-zero value of the condensate, the density of the quark modes
in infinite volume does not vanish at the origin. Conversely a 
non-zero density implies that the symmetry is broken. 

The mode number of the Dirac operator
\be
  \nu(\Lambda,m)=V \int_{-\Lambda}^{\Lambda} {\rm d}\lambda\,\rho(\lambda,m),
\ee
corresponds also to the average number of eigenmodes of the massive Hermitean 
operator $D^{\dagger}D+m^2$ with eigenvalues $\alpha\leq M^2 = \Lambda^2+m^2$.
It is a renormalisation-group invariant quantity as it stands. Its 
(normalized) discrete derivative 
\be\label{eq:discD}
{\tilde\rho}(\Lambda_1,\Lambda_2,m)  =  
\frac{\pi}{2 V} \frac{\nu(\Lambda_2)-\nu(\Lambda_1)}
{\Lambda_2 - \Lambda_1}\;  
\ee
carries the same information as $\rho(\lambda,m)$, but this {\it effective
spectral density} is a more convenient quantity to consider in practice
on the lattice. 

\subsection{Mode number on the lattice}
We discretize two-flavour QCD with the Wilson plaquette 
action for the gauge field, and O$(a)$-improved Wilson
action for the doublet of mass-degenerate 
quarks~\cite{Sheikholeslami:1985ij,Luscher:1996sc}, see appendix 
\ref{app:A} for more details. The mode number\footnote{We use the same
notation for lattice and continuum quantities, since any ambiguity is 
resolved from the context. As usual the continuum limit value of a
renormalised lattice quantity, identified with the subscript ${\rm R}$, is the one to 
be identified with its continuum counterpart.} $\nu(\Lambda,m)$ 
is defined as the average number  of eigenmodes of the massive 
Hermitean O$(a)$-improved Wilson-Dirac operator $D^\dagger_m D_m$ with eigenvalues 
$\alpha\leq M^2$. In the continuum limit this definition 
converges to the universal one~\cite{Giusti:2008vb} 
\be\label{eq:nuR}
\nu_{\R}(\Lambda_{\R},m_{\R}) = 
\nu(\Lambda,m)
\ee
provided $m_{\R}$ is defined as in Eq.~(\ref{eq:mR}), 
and $\Lambda_{\R}$ as 
\be
\Lambda_{\R}=\sqrt{M_{\R}^2-m_{\R}^2}\; , \qquad 
M_{\R} = Z^{-1}_{\rm P} (1 + {\bar b}_\mu\, a m)\, M\; .
\ee
The counter-term proportional to ${\bar b}_\mu$ ensures that at finite 
lattice spacing $\nu_{\R}(M_{\R},m_{\R})$ is an O$(a)$-improved 
quantity. This improvement coefficient has been computed in 
Ref.~\cite{Giusti:2008vb}, and its values for the inverse couplings 
$\beta$ considered
in this paper are given in Table~\ref{tab:impr}.

For Wilson fermions chiral symmetry is violated at finite lattice spacing. 
As a consequence the fine details of the spectrum of the Wilson--Dirac 
operator near the threshold $\Lambda_{\R}=0$ is not protected from large lattice 
effects~\cite{DelDebbio:2005qa,Damgaard:2010cz,Splittorff:2012gp}. While this 
region may be of interest for studying the peculiar 
details of those fermions, it is easier to extract universal information 
about the continuum theory  far away from it. In this respect
the effective spectral density in Eq.~(\ref{eq:discD}) is a good quantity to 
consider on the lattice to extract the value of the chiral 
condensate\footnote{Once the renormalisability
of the spectral density is proven, a generic finite integral 
of $\rho(\lambda,m)$ can be used to measure the condensate, see 
Ref.~\cite{Giusti:2007cn} for a different choice.} . 

\section{Numerical setup}
The CLS community\footnote{https://wiki-zeuthen.desy.de/CLS/CLS.} and the ALPHA 
Collaboration have generated 
the gauge configurations of the two-flavour
QCD with the $O(a)$-improved Wilson action by using the 
MP-HMC (lattices A5, B6, G8, N6 and O7) and the DD-HMC (all other lattices)
algorithms as implemented in 
Refs.~\cite{Marinkovic:2010eg,Luscher:DD-HMC}.
The primary observables that we 
have computed are the two-point functions of bilinear operators 
in Eq.~(\ref{eq:2pt}), and the mode number $\nu(\Lambda,m)$. 
The former were already computed by the ALPHA Collaboration, see 
Appendix~\ref{app:mmpif} and Refs.~\cite{Fritzsch:2012wq,alphalec} for 
more details.
\begin{table}
\small
\begin{center}
\setlength{\tabcolsep}{.10pc}
\begin{tabular}{@{\extracolsep{0.4cm}}cccccccccc}
\hline
id &$L/a$&$\beta$&$\kappa$&MDU&$m_{\R}$[MeV] &$F_\pi$[MeV]&$M_\pi$[MeV]&$M_\pi L$&$a$[fm]\\
\hline
A3  &$32$&$5.2$&$0.13580$ &$7040$ &$37.4(9)$ &$120.8(7)$& $496(6)$   & $6.0$  & 0.0749(8)\\     
A4  &$32$&     &$0.13590$ &$7920$ &$22.8(6)$ &$110.7(6)$&$386(5)$   & $4.7$  & \\
A5  &$32$&     &$0.13594$ &$1980$ &$16.8(4)$ &$106.0(6)$&$333(5)$   & $4.0$  & \\
B6  &$48$&     &$0.13597$ &$1200$ &$12.2(3)$ &$102.3(5)$&$283(4)$   & $5.2$  & \\
\hline
E5  &$32$&$5.3$&$0.13625$ &$8832$ &$32.0(8)$ &$115.2(6)$&$440(5)$   & $4.7$ & 0.0652(6)\\
F6  &$48$&     &$0.13635$ &$4000$ &$16.5(4)$ &$105.3(6)$&$314(3)$   & $5.0$ & \\
F7  &$48$&     &$0.13638$ &$3600$ &$12.0(3)$ &$100.9(4)$&$268(3)$   & $4.3$ & \\
G8  &$64$&     &$0.136417$&$1680$ &$6.1(2)$   &$95.8(4)$&$193(2)$& $4.1$ & \\
\hline
N5  &$48$&$5.5$&$0.13660$ &$3840$ &$34.8(8)$ &$115.1(7)$&$443(4)$   & $5.2$ & 0.0483(4)\\
N6  &$48$&     &$0.13667$ &$7680$ &$20.9(5)$ &$105.8(5)$&$342(3)$   & $4.0$ & \\
O7  &$64$&     &$0.13671$ &$3800$ &$12.9(3)$ &$101.2(4)$&$269(3)$   & $4.2$ & \\
\hline
\end{tabular}
\end{center}
\caption{\label{tab:ens} Overview of the ensembles and statistics used in this
study. We give the label, the spatial extent of the lattice,
$\beta=6/g_0^2$, the hopping parameter $\kappa$ for the quark fields, 
the number of molecular dynamics units (MDU), the quark mass $m_{\R}$
renormalized in the $\msbar$ scheme at $\mu=2$~GeV, 
the pion mass $M_\pi$ and its decay constant $F_\pi$, the product $M_\pi L$, and the (updated) value of
the lattice spacing determined as in~\cite{Fritzsch:2012wq} 
(see also \cite{Marinkovic:2011pa}).}
\end{table}

\subsection{Computation of the mode number} 
The stochastic computation of the mode number has been carried out 
as in Ref.~\cite{Giusti:2008vb}. A numerical approximation 
of the  orthogonal projector $\PM$ to the subspace spanned 
by the eigenmodes of $D_m^\dagger D_m$ with eigenvalues 
$\alpha\leq M^2$ is computed as
\be
  \PM\simeq h(\Xop)^4,
  \qquad
  \Xop=
  1-\frac{2 M_*^2}{D_m^\dagger D_m+ M_*^2}\; .
\ee
where $M/M_\star=0.96334$. The function $h(x)$ is an 
approximation to the step function $\theta(-x)$ by a 
minmax polynomial of degree $n=32$ in the range 
$-1\leq x \leq 1$, see Ref.~\cite{Giusti:2008vb} for 
more details. This choice, together with the value of 
$M_\star$ given, guarantees a systematic 
error well below our statistical errors.
The mode number is then computed as 
\be
  \nu(M,m)=\langle{\cal O}_N\rangle,
  \qquad
  {\cal O}_N={1\over N}\sum_{k=1}^N\left(\eta_k,\PM\eta_k\right),
\ee
where we have added to the theory a set of pseudo-fermion fields 
$\eta_1,\ldots,\eta_N$ with Gaussian action.  In the course of a numerical 
simulation, one such field ($N=1)$ for each gauge-field configuration is
generated randomly, and the mode number is estimated in the usual way 
by averaging the observable ${\cal O}_N$ over the generated
ensemble of fields. The mode number is an extensive quantity, and 
at fixed $N$ and for a given statistics, the relative statistical error 
of the calculated mode number is therefore expected to decrease like 
$V^{-1/2}$.

\subsection{Ensembles generated}
The details of the lattices are listed in Tables~\ref{tab:ens} and 
\ref{tab:ens2}. All of them have a size of $2L \times L^3$, and the 
spatial dimensions are always large enough so that $M_\pi L\geq4$. 
The three values of the coupling constant $\beta=5.2,\,5.3,\,5.5$ 
correspond to lattice spacings of $\;a=0.075,\, 0.065,\, 0.048$\,fm  
respectively, which have been fixed from $F_K$ by supplementing 
the theory with a 
quenched ``strange'' quark~\cite{Fritzsch:2012wq}. The pion 
masses range from $190$ MeV to $500$ MeV. To explicitly check for 
finite-size effects in the mode number we have generated an additional set of 
lattices (D5) with the same spacing and quark mass as E5, but with 
a smaller lattice volume $48\times 24^3$.

\begin{table}
\small
\begin{center}
\begin{tabular}{@{\extracolsep{0.0cm}}cc|cc|cc|c}
\hline
id & $R_{\rm act}$ & 
$R_{\rm act}\tau_{\rm int}(M_{\pi})$&$R_{\rm act} n_{\rm it}(M_{\pi})$ & 
$R_{\rm act}\tau_{\rm int}(\nu)$   &$R_{\rm act} n_{\rm it} (\nu)$ &
$R_{\rm act}\tau_{\rm exp}$\\
\hline
A3 &$0.37$& 7 & 2.96 &               & \phantom{1}47.36 & 40\\
A4 &$0.37$& 5 & 2.96 &               & \phantom{1}53.28 &   \\
A5 & 1    & 5 & 4.00 & 3             & \phantom{1}36.00 & \\
B6 & 1    & 6 & 2.00 &               & \phantom{1}24.00 & \\
\hline
E5 &$0.37$& 9 & 5.92 & 6             & \phantom{1}35.52 & 55\\
F6 &$0.37$& 8 & 2.96 &               & \phantom{1}29.60 &   \\
F7 &$0.37$& 7 & 2.96 &               & \phantom{1}26.64 &   \\
G8 &  1   & 8 & 2.00 &               & 24--48 &   \\
\hline
N5 &$0.44$&30 & 3.52 & 11            & \phantom{1}28.16 & 100\\
N6 & 1    &10 & 4.00 &               & 128 &    \\
O7 & 1    &15 & 4.00 &               & \phantom{1}76 &    \\
\hline
\end{tabular}
\end{center}
\caption{\label{tab:ens2} The integrated autocorrelation time
$\tau_{\rm int}$ of the pion mass and of the mode number, multiplied 
by the fraction of active links in the HMC $R_{\rm act}$,
is given in units of MDU. The parameters $\tau_{\rm int}$ have a typical
error of $25$--$35\%$. The number $n_{\rm it}$ of MDUs skipped between two 
consecutive measurements of the two-point functions and of the mode 
number is also reported. The value of $\tau_{\rm exp}$ of the Markov chain 
given in the last column is taken from Ref.~\cite{Bruno:2014ova}. The value 
of $R_{\rm act}\tau_{\rm int}(\nu)$
for N5 is a conservative estimate from the one of E5 and a
scaling proprtional to $\tau_{\rm exp}$.}
\end{table}
The autocorrelation times of the two-point functions and of the mode 
number are reported in Table~\ref{tab:ens2}. For the lattice E5 we 
have computed $\tau_{\rm int}(\nu)$ for three values of $aM$ 
corresponding to $\Lambda_{\R} = 30,\, 40$ and $86$~MeV, and no significative
difference was observed. We thus space the 
measurements to give time to the mode number to decorrelate, while we bin 
properly the (cheaper) measurements of the two-point functions. To measure $\nu$,
the number of configurations to be processed is chosen so that the statistical
error of the effective spectral density receives roughly equally-sized contributions from 
the scale and the mode number. To ensure a proper Monte Carlo sampling, a 
minimum of 50 configurations is processed in any case.

The value of $\tau_{\rm exp}$ of the Markov chain, defined as in 
Ref.~\cite{Fritzsch:2012wq}, is taken from~\cite{Bruno:2014ova}.
It gets significantly longer towards finer lattice spacings. 
For the ensembles
where $n_{\rm it}<\tau_{\rm exp}$, we estimate the contributions of the tails in the 
autocorrelation functions of the observables as described in Ref.~\cite{Schaefer:2010hu}.
When needed, we take them into account to have a more conservative error estimate.

\section{A first look into the numerical results}\label{sec:firstlook}
We have computed the mode number $\nu$ for 
nine values\footnote{If not explicitly stated, the scheme- and 
scale-dependent quantities such as $\Sigma$, $m_{\R}$, $\Lambda_{\R}$ and 
$\wS$ are  renormalized in the $\msbar$ scheme at 
$\mu=2$~GeV.} of $\Lambda_{\R}$ in the range $20$--$120$~MeV 
with a statistical accuracy of a few percent on all lattices 
listed in Table~\ref{tab:ens}. Four larger
values of $\Lambda_{\R}$ in the range $150$--$500$~MeV have 
been also analysed for the ensemble E5. The results are collected 
in Tables~\ref{tab:lambda}--\ref{tab:lambda3}  of the 
Appendix~\ref{sec:parms}. 
\begin{figure}
\begin{minipage}{0.35\textwidth}
\includegraphics[width=7.0 cm,angle=0]{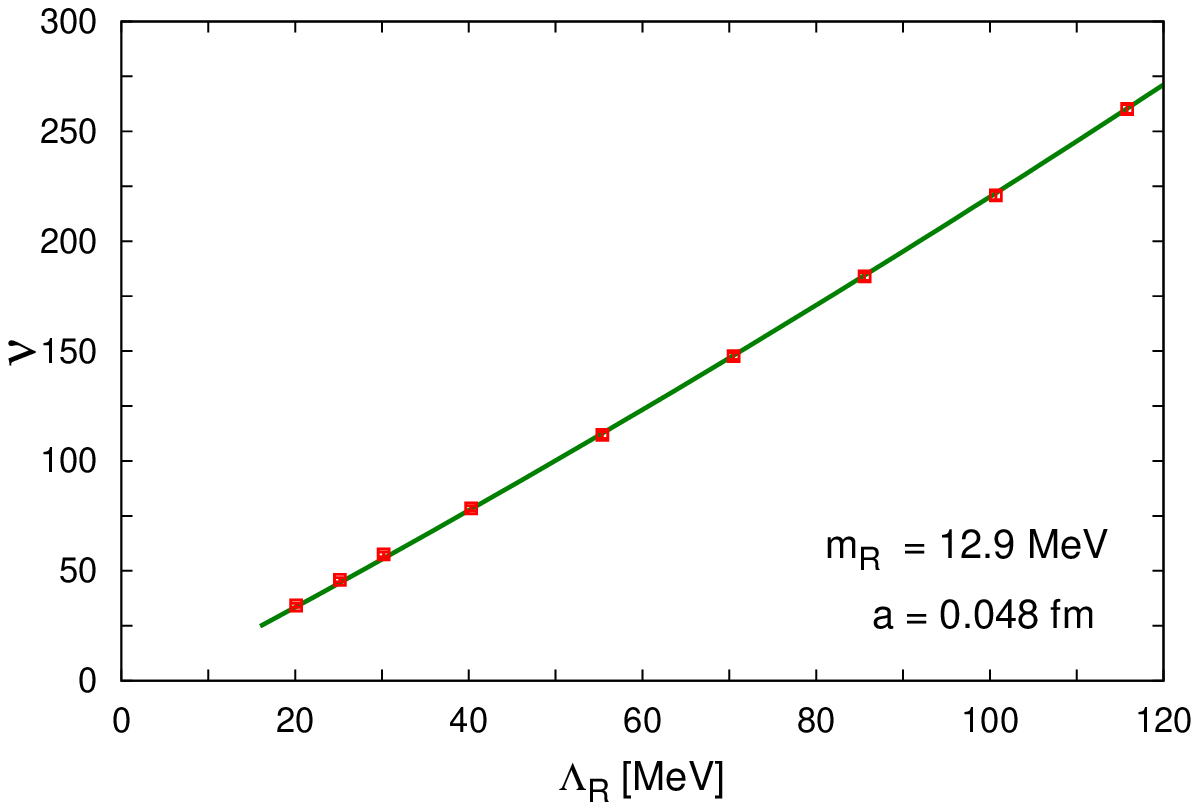}
\end{minipage}
\hspace{20mm}
\begin{minipage}{0.35\textwidth}
\includegraphics[width=7.0 cm,angle=0]{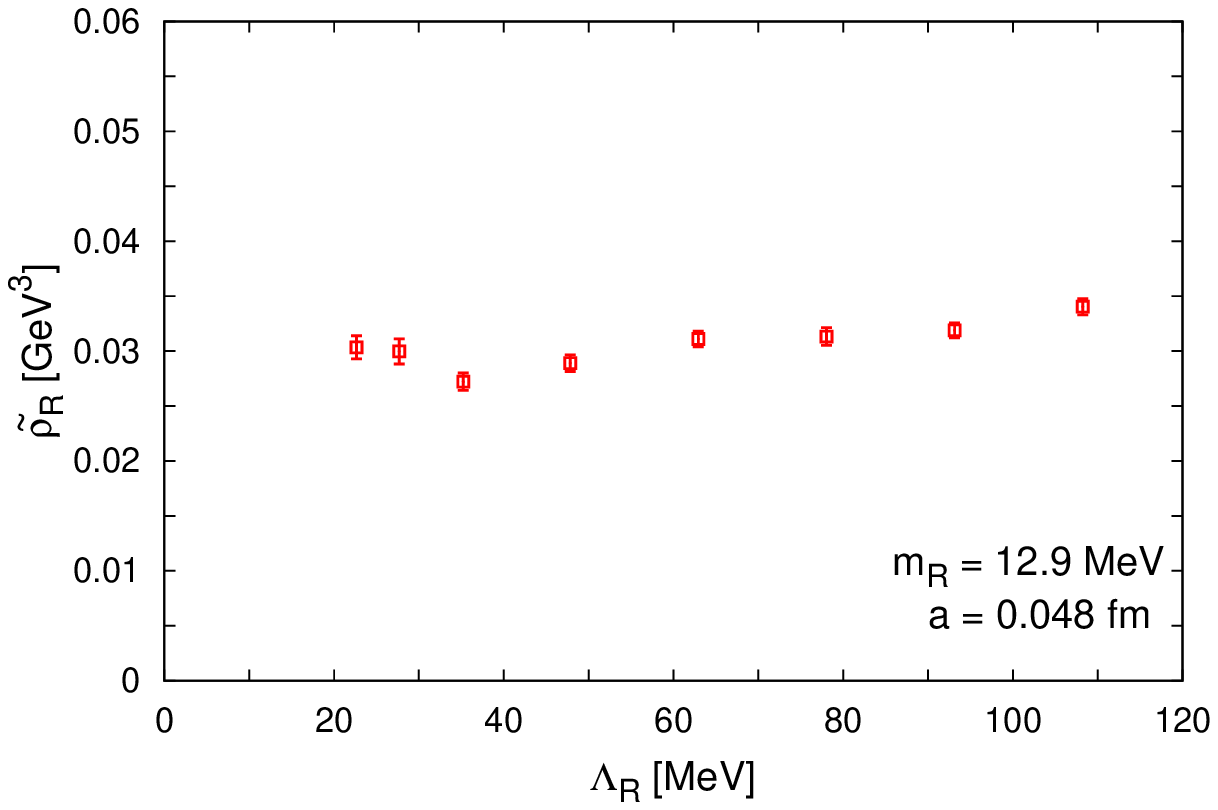}
\end{minipage}
\caption{Left: the mode number $\nu$ as a function of $\Lambda_{\R}$
for the ensemble O7. A quadratic fit of the data gives  
$\nu= -9.0(13) + 2.07(7)\Lambda_{\R} + 0.0022(4)\Lambda_{\R}^2$. Right:
the effective spectral density $\wS$ as defined in Eq.~(\ref{eq:discD}) 
for the same ensemble as a function 
of $\Lambda_{\R}=(\Lambda_{1,\R}+\Lambda_{2,\R})/2$. Since we are interested 
in the $\Lambda_{\R}$-dependence only, the errors in this plot do not include 
those of the lattice spacing and of $Z_{\rm P}$. 
The errors from $Z_{\rm A}$  and $m_{\R}$ appear to be invisible in the figure.}
\label{fig:firstlook-O7}
\end{figure}

In Figure~\ref{fig:firstlook-O7} we
show $\nu$ as a function of $\Lambda_{\R}$ for the lattice O7, 
corresponding to the smallest reference quark mass (see below) at the smallest 
lattice spacing. On all other lattices an analogous qualitative 
behaviour is observed. 
The mode number is a nearly linear function 
in $\Lambda_{\R}$ up to approximatively $100$--$150$~MeV. A clear departure from
linearity is observed for $\Lambda_{\R} > 200$~MeV on the 
lattice E5.  At the percent precision, however, the data show 
statistically significant deviations from the linear behavior already
below $100$ MeV. To guide the eye, a quadratic fit in $\Lambda_{\R}$ is shown in 
Figure~\ref{fig:firstlook-O7}, and the values of the 
coefficients are given in the caption. The bulk of $\nu$
is given by the linear term, while the constant and the quadratic 
term represent $O(10\%)$ corrections in the fitted range. The nearly
linear behaviour of the mode number is manifest on the right plot of 
Figure~\ref{fig:firstlook-O7}, where its discrete derivative, 
defined as in Eq.~(\ref{eq:discD}) for each couple of 
consecutive values of $\Lambda_{\R}$, is shown as a function of 
$\Lambda_{\R}=(\Lambda_{1,\R}+\Lambda_{2,\R})/2$. Since it is not affected by threshold 
effects, the effective spectral density $\wS$ is the primary observable
we focus on in the next sections.

\begin{figure}
\begin{center}
\includegraphics[width=7.0 cm,angle=0]{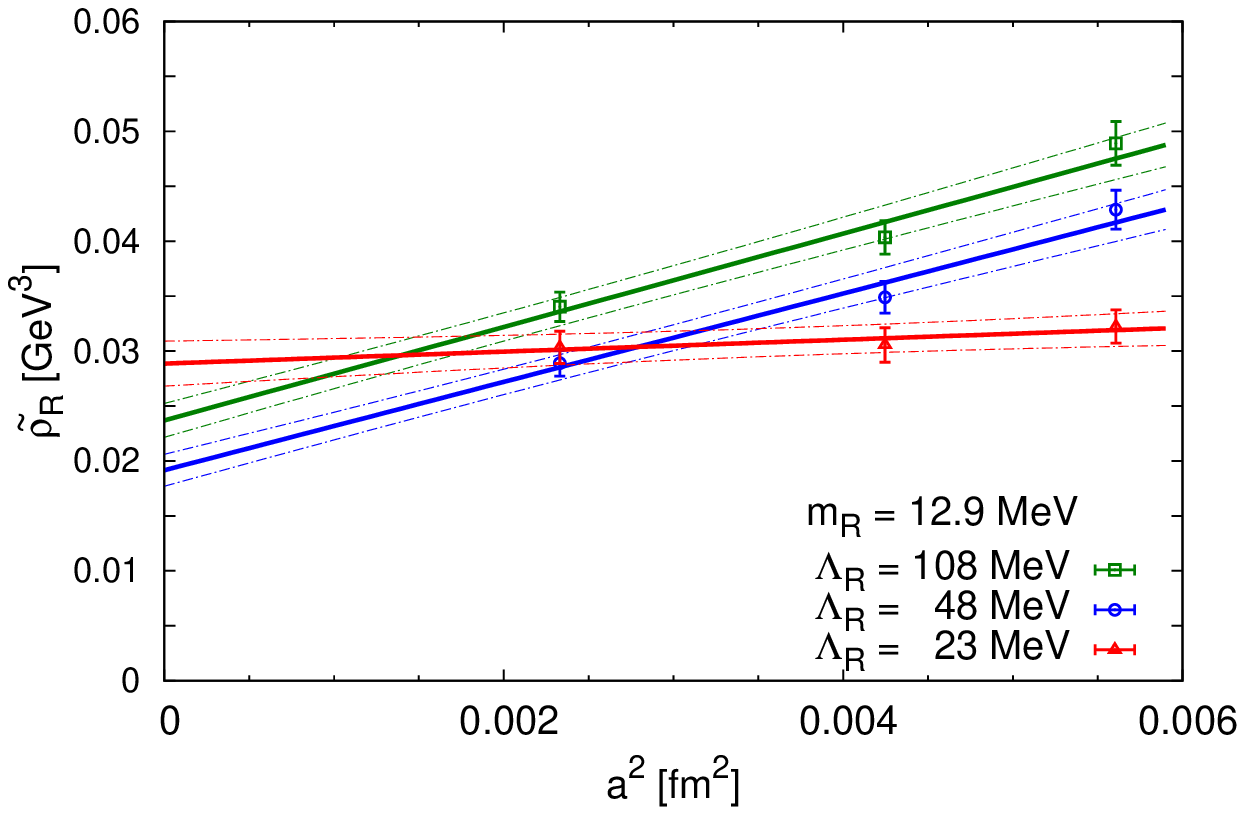}
\includegraphics[width=7.0 cm,angle=0]{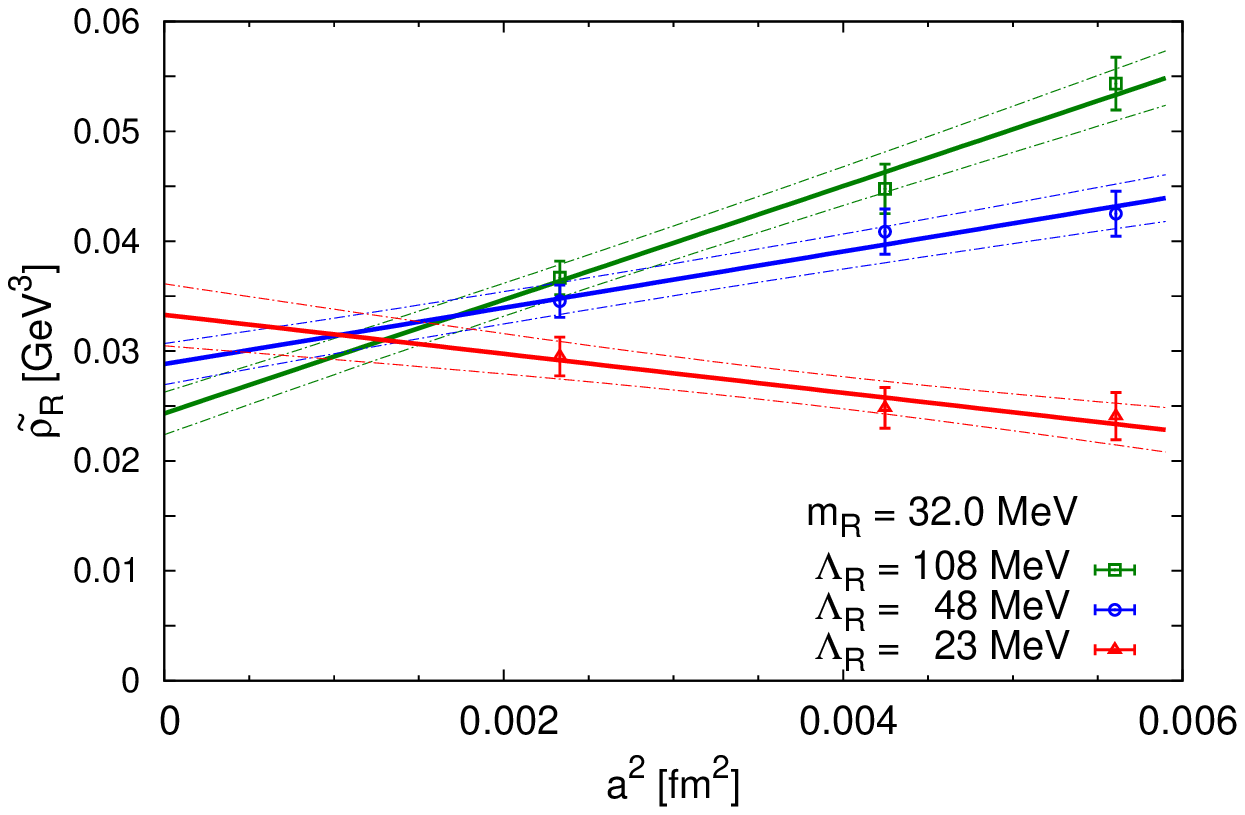}
\caption{Effective spectral density $\wS$ vs.~the lattice spacing squared 
for the lightest (left hand side) and the heaviest reference quark mass 
$m_{\R}$ (right hand side), and for the lightest, an intermediate, and the heaviest cutoff $\Lambda_{\R}$ in both panels. 
In general, the data are well described by a linear fit in $a^2$, which suggests 
that, within our statistical errors, we are in the asymptotic regime of Symanzik 
effective theory. 
As evident from the figures, there are competing (positive and negative) discretization 
effects, which can approximately compensate for each other in specific domains of parameter 
space. 
}
\label{fig:contlim}
\end{center}
\end{figure}

\subsection{Continuum-limit extrapolation}
\begin{figure}
\begin{minipage}{0.35\textwidth}
\includegraphics[width=7.0 cm,angle=0]{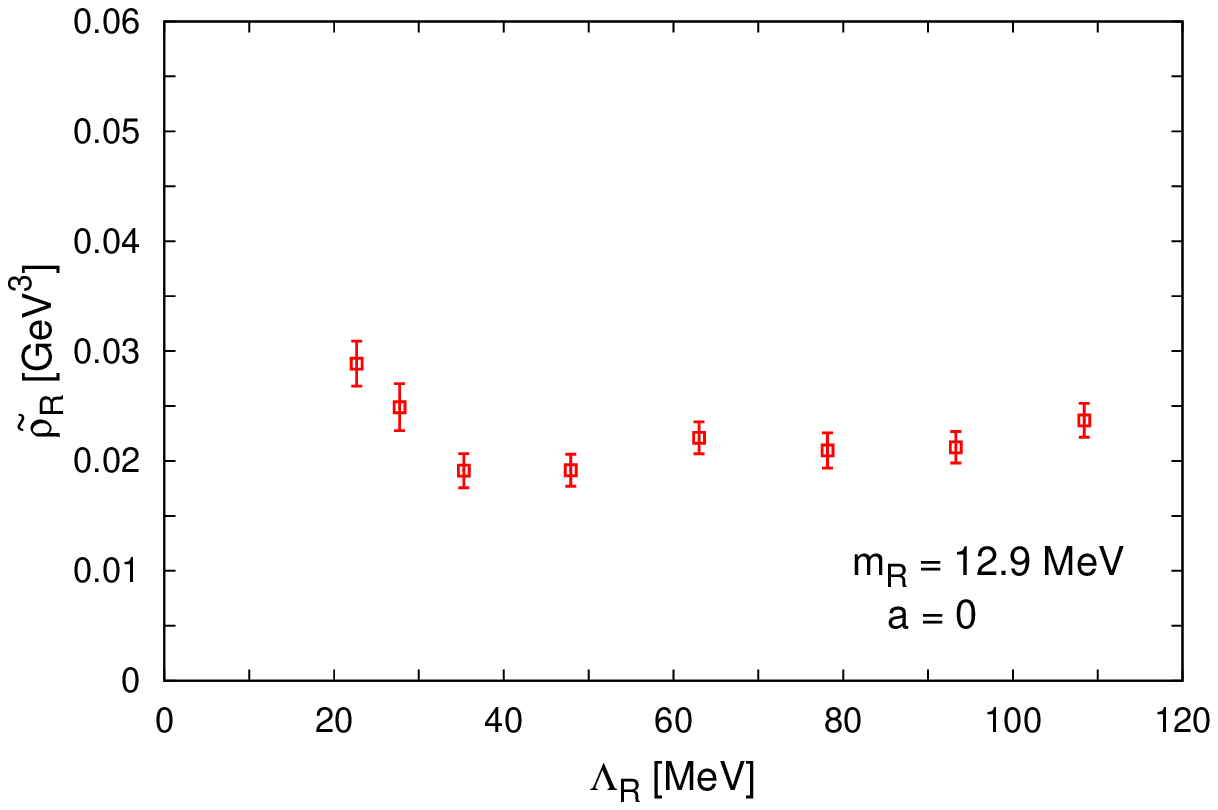}
\end{minipage}
\hspace{20mm}
\begin{minipage}{0.35\textwidth}
\includegraphics[width=7.0 cm,angle=0]{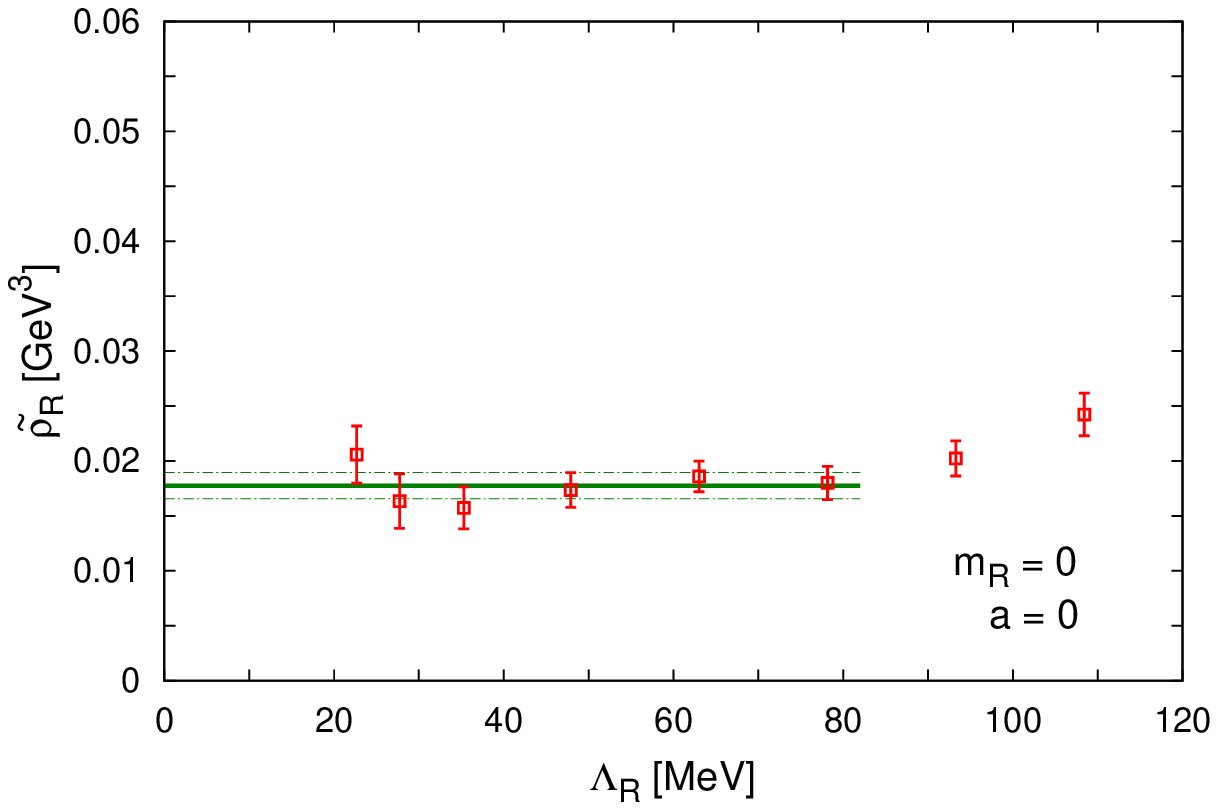}

\end{minipage}
\caption{Effective spectral density $\wS$ in the continuum limit at the 
smallest reference quark 
mass $m_{\R}=12.9$~MeV (left), and in the chiral limit (right). 
Note the flat dependence on $\Lambda_{\R}$ which agrees with the expectation 
from NLO ChPT. The results of the fit to a constant is also shown on the 
right plot.
}
\label{fig:firstlook-contlim}
\end{figure}

In general for $\wS$ we observe quite a flat behaviour in 
$\Lambda_{\R}$ toward finer lattice spacings and light quark 
masses, similar to the one shown in Figure~\ref{fig:firstlook-O7}. Because 
the action and 
the mode number are $O(a)$-improved, the Symanzik effective theory analysis
predicts that discretization errors start 
at $O(a^2)$. In order to remove them, at every lattice spacing 
we match three quark mass values ($m_{\R}=12.9$, $20.9$, $32.0$~MeV) 
by interpolating $\wS$ linearly in $m_{\R}$ (see next section for more details).
The values of $\wS$ show mild discretization effects at light $m_{\R}$
and $\Lambda_{\R}$, while they differ up to $15\,\%$ per linear dimension 
among the three lattice spacings toward larger $\Lambda_{\R}$. Within the  
statistical errors all data sets are compatible with a linear dependence 
in $a^2$, and we thus independently extrapolate each triplet of points
to the continuum limit accordingly. We show six of those extrapolations 
in Figure \ref{fig:contlim}, considering the lightest and the heaviest reference 
quark masses for
the lightest, an intermediate, and the heaviest cutoff $\Lambda_{\R}$.  
The difference between the values of 
$\wS$ at the finest lattice spacing and the continuum-extrapolated 
ones is within the statistical errors for light $m_{\R}$ and $\Lambda_{\R}$, and
it remains within few standard deviations toward larger values of $m_{\R}$ 
and $\Lambda_{\R}$. This fact makes us confident that the extrapolation removes 
the cutoff effects within the errors quoted.\\
The results
for $\wS$ at $m_{\R}=12.9$~MeV in the continuum limit are shown as a function
of $\Lambda_{\R}$ in the left plot of Figure~\ref{fig:firstlook-contlim}. 
A similar $\Lambda_{\R}$-dependence is observed at the two other 
reference masses. 
It is worth noting that no assumption on the presence of 
spontaneous symmetry breaking was needed so far.
These results, however, point to the fact that the spectral density 
of the Dirac operator in two-flavour QCD is (almost) constant 
in $\Lambda_{\R}$ near the origin at small quark masses. This is consistent 
with the expectation from the Banks--Casher relation in presence of 
spontaneous symmetry breaking. In this case next-to-leading (NLO) ChPT 
indeed predicts
\be\label{eq:eqChPTtext}
\wS^{\rm nlo} =  
\Sigma \Big\{ 1 + \frac{m_{\R} \Sigma}{(4\pi)^2 F^4} \Big[3\, \bar l_6 + 1 - 
\ln(2) - 3 \ln\Big(\frac{\Sigma m_{\R}}{F^2 \bar\mu^2}\Big) + 
\tilde g_\nu\left(\frac{\Lambda_{1,\R}}{m_{\R}},\frac{\Lambda_{2,\R}}{m_{\R}}\right) 
\Big]\Big\}\; ,
\ee
i.e.~an almost flat function in (small) $\Lambda_{\R}$ at (small) finite quark 
masses, see Appendix~\ref{app:modenumber:chpt} for 
unexplained notation. At fixed quark mass the $\Lambda_{\R}$-dependence 
of $\wS^{\rm nlo}$ in Eq.~(\ref{eq:eqChPTtext}) is 
parameter-free once the pion mass and decay constant are measured. 

\subsection{Chiral limit}
The extrapolation to the chiral limit requires an assumption 
on how the effective spectral density $\wS$ behaves when 
$m_{\R} \rightarrow 0$. In this respect it is interesting to note that 
the NLO function in Eq.~(\ref{eq:eqChPTtext}) goes linearly in 
$m_{\R}$ near the chiral limit since there are no chiral logarithms 
at fixed $\Lambda_{\R}$, see Appendix \ref{app:modenumber:chpt}. 
A fit of the data to Eq.~(\ref{eq:eqChPTtext}) shows that the data are 
compatible with that NLO formula. A prediction of NLO ChPT in 
the two-flavour theory
is that in the chiral limit $\wS^{\rm nlo}=\Sigma$ also at non-zero 
$\Lambda_{\R}$, since all NLO corrections in Eq.~(\ref{eq:eqChPTtext})
vanish~\cite{Smilga:1993in}. To check
for this property we extrapolate $\wS$ with 
Eq.~(\ref{eq:NLO-min-extended}), which is a generalization of 
Eq.~(\ref{eq:eqChPTtext}) see below, and we obtain 
the results shown in the right plot of 
Figure~\ref{fig:firstlook-contlim} with a $\chi^2/\rm{dof}=16.4/14$. 
Within errors 
the $\Lambda_{\R}$-dependence is clearly compatible with a constant up to 
$\approx 80$~MeV. Moreover the difference between the values of 
$\wS$ in the chiral limit and those at $m_{\R}=12.9$~MeV is of the order 
of the statistical error, i.e. the extrapolation is very mild.
A fit to a constant of the data gives $\Sigma^{1/3}=261(6)$~MeV.

As in any numerical computation, the chiral limit inevitably requires an 
extrapolation of the results with a pre-defined functional form. The 
distinctive feature of spontaneous symmetry breaking, however, is that the 
behaviour of $\wS$ near the origin is predicted by ChPT, and its extrapolated
value has to agree with the one of $M_\pi^2 F_\pi^2/(2 m_{\R})$. We have 
thus complemented our computations with those for $m_{\R}$, $M_\pi$ and $F_\pi$, and 
extrapolated the above mentioned ratio to the chiral limit as prescribed by ChPT, see
Appendix \ref{app:mmpif} and Ref.~\cite{Engel:2014cka}. 
We obtain $\Sigma^{1/3}_{\rm GMOR}= 263(3)(4)$~MeV, where
the first error is statistical and the second is systematic, in excellent 
agreement with the value quoted above. These results show that the spectral 
density at the origin has a non-zero value in the chiral limit. In the rest of this 
paper we assume this conclusion, 
and we apply standard field theory arguments to remove with confidence the (small) contributions 
in the raw data due to the discretization effects, the finite quark mass and finite 
$\Lambda_{\R}$.

\section{Detailed discussion of numerical results\label{sec:glbfit}}
We have analysed the numerical results for the effective 
spectral density $\wS$ following two different fitting strategies.
In the first one, the main results of which are reported in the 
previous section, we have extrapolated the results at 
fixed kinematics $(\Lambda_{\R},m_{\R})$ to the continuum limit
independently. The results of this analysis call
for an alternative strategy to extract the chiral condensate 
which uses ChPT from the starting point, i.e. based on fitting 
the data in all three directions $(\Lambda_{\R},m_{\R},a)$ at the 
same time. This procedure reduces the number of fit parameters, 
allows us to include all generated data in the fit, and 
avoids the need for an interpolation in the quark mass. 
It is important to stress that also in this case ChPT is used to 
remove only (small) higher order corrections in the spectral 
density. The details of these fits are reported in the 
next two sub-sections.

\subsection{Continuum limit fit\label{sec:seccont}}
In the first strategy outlined in Section \ref{sec:firstlook} 
we start by interpolating the data in the
quark mass at fixed $\Lambda_{\R}$ and $a$. We choose three reference 
values ($m_{\R}=12.9$, $20.9$, $32.0$~MeV) which are within the range of 
simulated quark masses at all $\beta$ values, and they are as close as 
possible to the 
values at the finest lattice spacing. Most of the data sets look 
perfectly linear in $m$ in the vicinity of the interpolation points, 
with small deviations only for
simultaneous coarse lattices, light $\Lambda_{\R}$'s and towards heavy quark 
masses (see Figure~\ref{fig:globalfit1}). In all cases, however, the 
systematic error associated with the linear interpolation is negligible 
with respect to the statistical one. The interpolation and all following fits 
are performed using the jackknife technique to take into account the 
correlation of the data. 

At fixed $(\Lambda_{\R},m_{\R})$, each data set is well fitted by a linear function 
in $a^2$, see Figure~\ref{fig:contlim}, a fact which supports the assumption
of being in the Symanzik asymptotic regime within the errors 
quoted\footnote{A detailed analysis of discretization effects in the spectral density
is beyond the scope of this paper. For completeness we report the results of 
these fits in Appendix D for the interested readers.}. 
Once extrapolated to the continuum 
limit, we fit the effective spectral density with the functional form 
\be\label{eq:NLO-min-extended}
\wS = c_0(\Lambda_{\R}) +
m_{\R} \Big[c_1 + c_2 
\Big(-3\ln\left(\frac{m_{\R}}{\bar\mu}\right)
+\tilde 
g_\nu\left(\frac{\Lambda_{\R,1}}{m_{\R}},\frac{\Lambda_{\R,2}}{m_{\R}}\right)\Big)\Big] \; , 
\ee
which rests on NLO ChPT, but it is capable of accounting for 
$O(\Lambda^2)$ effects. The latter are expected to be 
the dominant higher order effects in ChPT in this range of parameters. 
Within the given accuracy, $c_0(\Lambda)$ is consistent with a plateau 
behaviour in the range 
$20\leq\Lambda_{\R}\leq 80$ MeV, see right plot of Figure~\ref{fig:firstlook-contlim}. 
By fitting $c_0(\Lambda_{\R})$ to a constant in this range,
we obtain $\Sigma^{1/3}=261(6)~$~MeV. If we include also a $\Lambda_{\R}^2$ term in the 
fit and consider the entire range $20\leq\Lambda_{\R}\leq120$ MeV  
we find $253(9)$~MeV, which differs from the previous result by roughly one standard 
deviation. At the level of our statistical errors of $O(10\%)$, the spectral 
density of the Dirac operator in the continuum and chiral limits is a constant 
function up to $\Lambda_{\R} \approx 80$~MeV.

\subsection{Combined fit\label{sec:seccombined}} 
 
\begin{figure}
\begin{center}
\includegraphics[width=7.0 cm,angle=0]{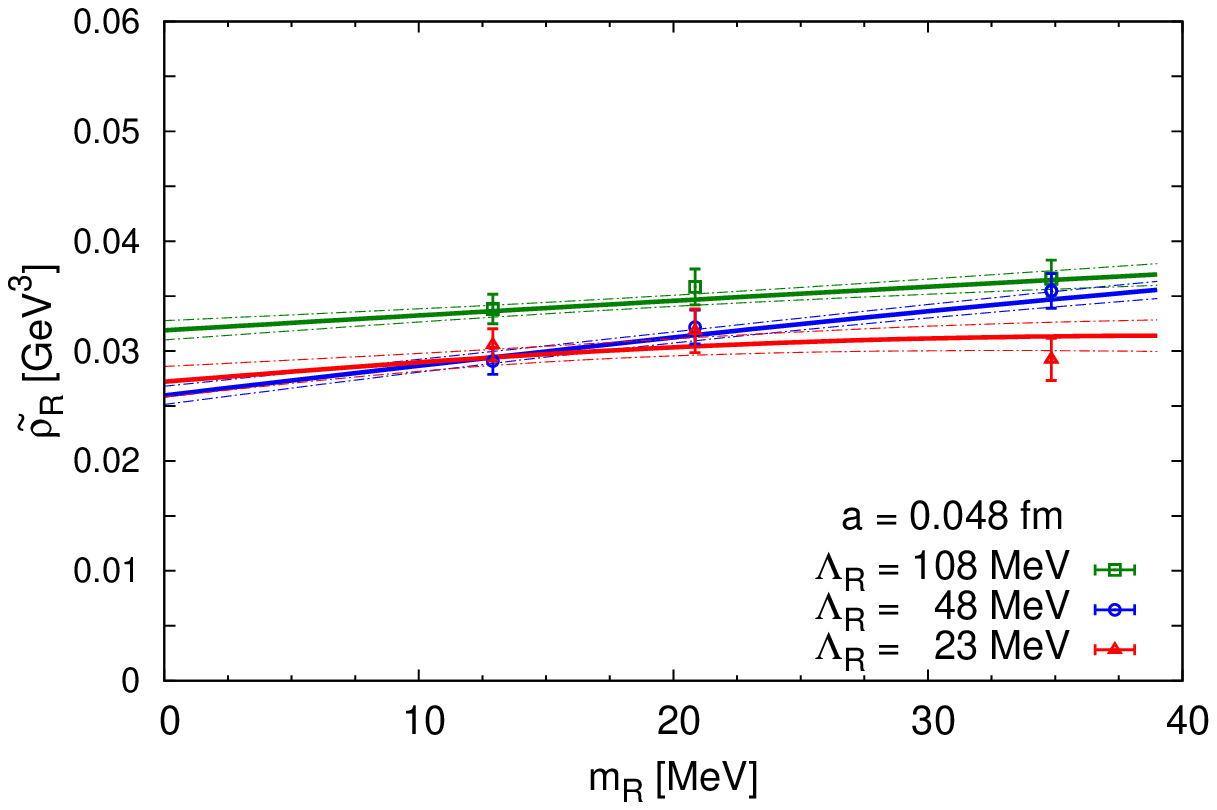}
\includegraphics[width=7.0 cm,angle=0]{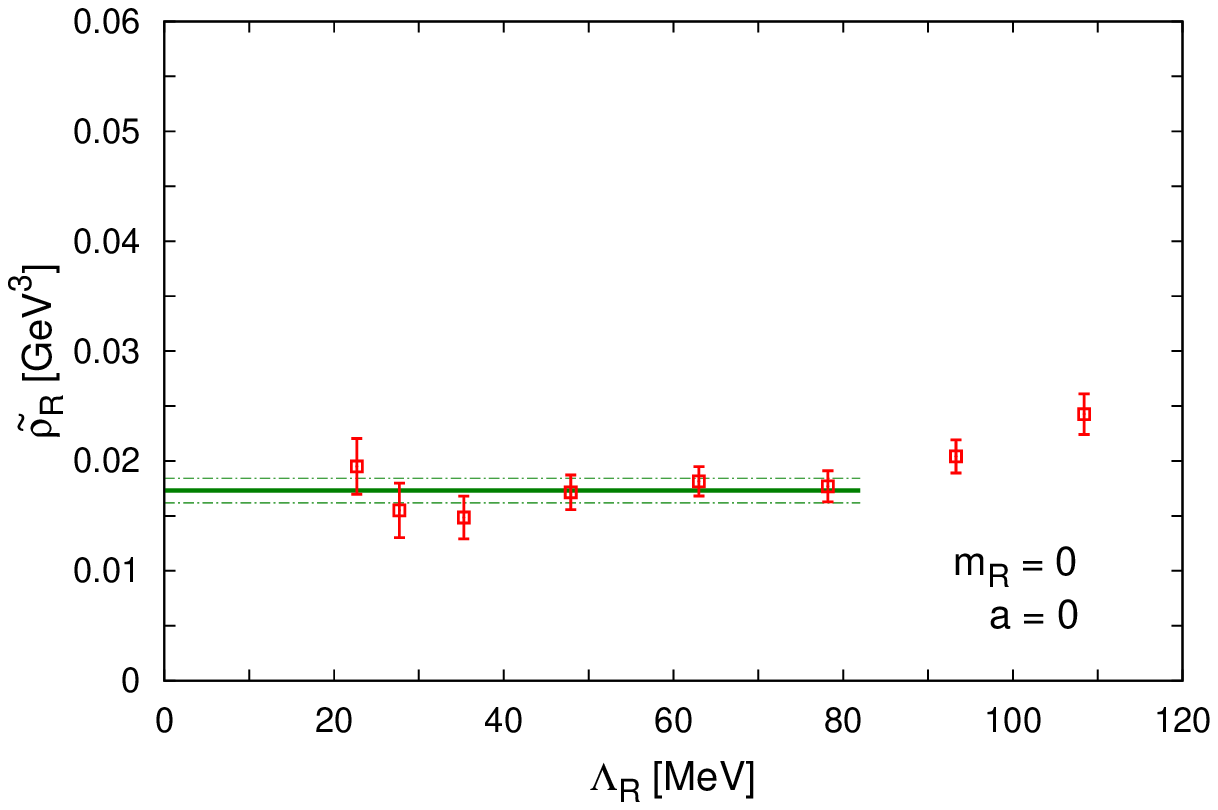}
\caption{Left: effective spectral density $\wS$ vs.~the quark mass $m_{\R}$ for the finer 
lattice spacings and 
three cutoffs $\Lambda_{\R}$ together with the combined fit to all data  
to Eq.~\eqref{eq:NLO-extended-reduced}. Right: effective spectral density $\wS$ vs.~the 
cutoff $\Lambda_{\R}$ in the continuum and chiral limits. The squares are the 
results for $c_{0,0}(\Lambda_{\R})$ of the fit to the function in 
Eq.~\eqref{eq:NLO-extended-reduced}, and the plateau fit shown gives the value for 
the chiral condensate.}
\label{fig:globalfit1}
\end{center}
\end{figure}
In this section we present an alternative strategy to extract the chiral condensate, 
based on fitting the data in all three directions $(\Lambda_{\R},m_{\R},a)$ at the same time. 
Compared to the first strategy, the shortcomings are that we cannot disentangle different
corrections as clearly and ChPT is used from the very beginning. We remark, however, that 
also in this case ChPT is used only to remove higher order corrections, while the bulk 
of the 
chiral condensate is still given through the Banks--Casher relation. The statistical analysis
is based on a double-elimination jackknife fit to take into account all errors and correlations 
(no fit of fitted quantities is needed). We start with the fit form 
\be\label{eq:NLO-extended}
\wS = c_0(\Lambda_{\R},a) +
m_{\R} \Big[c_1(\Lambda_{\R},a) + c_2 
\Big(-3\ln\left(\frac{m_{\R}}{\bar\mu}\right)
+\tilde 
g_\nu\left(\frac{\Lambda_{\R,1}}{m_{\R}},\frac{\Lambda_{\R,2}}{m_{\R}}\right)\Big)\Big]\; ,
\ee
where $\Lambda_{\R}=(\Lambda_{\R,1} + \Lambda_{\R,2})/2$, and we constrain the fit-parameters 
as suggested by NLO chiral and Symanzik effective theories. 
As already verified in the 
first strategy, the discretization effects obey an $a^2$-dependence in the range 
of parameters simulated. We thus constrain our fit parameters to 
obey \footnote{Note that this expression includes also the functional 
form of discretization effects predicted at NLO in the GSM regime of ChPT~\cite{Necco:2011vx}, 
see Appendices \ref{app:modenumber:chpt} and \ref{app:disc}.}
\be
c_0(\Lambda_{\R},a) = c_{0,0}(\Lambda_{\R}) + a^2 c_{0,1}(\Lambda_{\R})\; , \qquad
c_1(\Lambda_{\R},a) = c_{1,0}(\Lambda_{\R}) + a^2 c_{1,1}(\Lambda_{\R})\; . 
\ee 
The NLO ChPT predicts that $c_{0,0}(\Lambda_{\R})$ and 
$c_{1,0}(\Lambda_{\R})$ should both be constant. Allowing for the time being an arbitrary $\Lambda_{\R}$-dependence 
in the parameter $c_{0,0}(\Lambda_{\R})$, we arrive at the fit function 
\bea \label{eq:NLO-extended-reduced}
\hspace{-8.0cm} \wS & = & c_{0,0}(\Lambda_{\R}) + a^2 c_{0,1}(\Lambda_{\R})
+  m_{\R} \Big[c_{1,0} + a^2 c_{1,1}(\Lambda_{\R})
+ \nonumber\\ 
& &\qquad\qquad\qquad\qquad\qquad\qquad\quad  c_2 
\Big(-3\ln\left(\frac{m_{\R}}{\bar\mu}\right)
+\tilde 
g_\nu\Big(\frac{\Lambda_{\R,1}}{m_{\R}},\frac{\Lambda_{\R,2}}{m_{\R}}\Big)\Big)\Big] \;.
\eea 
The fit of the data is shown versus the quark mass 
in the left plot of Figure~\ref{fig:globalfit1} for the finer lattice spacings 
and three cutoffs $\Lambda_{\R}$'s. 
The resulting effective spectral density in the continuum and chiral limit, corresponding 
to $c_{0,0}(\Lambda_{\R})$, is shown in the right plot of Figure~\ref{fig:globalfit1}. The results 
are very well compatible 
with the ones determined in Section~\ref{sec:firstlook}. If we fix $c_{0,0}$ to a constant in the 
region $20\leq \Lambda_{\R} \leq 80$, we can extract the condensate to get 
$\Sigma^{1/3}=259(6)~$~MeV, which is well compatible with the one extracted in the 
previous strategy. 

To assess the stability of the fit we have amended the function with higher order 
terms of the form $\mathcal{O}(\Lambda_{\R}^2,\Lambda_{\R} m_{\R},m_{\R}^2)$. 
Note that when including $\Lambda_{\R}^2$ terms, we always consider the entire range 
$20\leq \Lambda_{\R} \leq120$ MeV. 
The coefficient of $\Lambda_{\R} m_{\R}$ is consistent with zero, 
while $m_{\R}^2$ and $\Lambda_{\R}^2$ effects are non-zero by 2 and 3 standard deviations
respectively and affect our final result systematically by roughly 1 standard deviation 
downwards. We remark, however, that in the truncated range $20\leq \Lambda_{\R} \leq80$ 
the data is perfectly compatible with a flat dependence on $\Lambda_{\R}$. We also investigated the 
effect of truncating the amount of data included in the fit. Cutting light $\Lambda_{\R}$ slightly 
improves 
the fit, while cutting heavy ones does not make a noteworthy difference. 
To check again whether all data obey well the assumed linear $a^2$-dependence, we 
perform also fits excluding the data at the coarsest lattices ($a=0.075$ fm)
with larger discretization effects (we kept 12 out of 32 data points at this 
lattice spacing). This does not improve the quality of the fit significantly, and it
gives $\Sigma^{1/3}=267(6)$~MeV which differs from the previous result by roughly one
standard deviation upwards. We remark, however, that the linear $a^2$-dependence has been 
checked and confirmed explicitly for each pair of $(\Lambda_{\R},m_{\R})$ in the first 
strategy. A further reduction of the number of fit parameters can actually be achieved by 
noting that $c_2$ is known 
in ChPT. One can rewrite it as a function of $m_\pi$ and $m$. We have also tried to fix 
$c_{0,1}(\Lambda_{\R})$ to a 
constant which is suggested from results of the several fits we have done 
(see Appendix~\ref{app:disc}). In either case we get results which are well compatible with 
the results quoted. 

For this strategy the best value of the chiral condensate is 
$\Sigma^{1/3}=259(6)$~MeV. It is extracted 
from the fit function Eq.~\eqref{eq:NLO-extended-reduced} where $c_{0,0}$ is fitted to a constant in the range 
$20\leq \Lambda_{\R} \leq 80$ MeV. 
This fit confirms that in the chiral and continuum limits the spectral density is a flat function
of $\Lambda_{\R}$ up to $\approx 80$~MeV at the level of our precision in the continuum limit of 
roughly $10\%$, and it can be parameterized by NLO ChPT.

We presented preliminary results of this study at only two lattice spacings 
in Ref.~\cite{Engel:2013rwa}. There we observed effects of $\mathcal{O}(\Lambda_{\R}^2)$ 
already for $\Lambda_{\R}\gtrsim50$ MeV, in particular for $a=0.065$~fm. Once the data
are extrapolated to the continuum limit, these effects are not visible anymore
up to $\Lambda_{\R}\approx 80$ MeV. In this respect it must be noted, however, that once the 
uncertainties in the scale and renormalisation constants are included, the final 
errors of the extrapolated results are significantly larger than those used  to study 
the $\Lambda_{\R}$-dependence at fixed lattice spacing. It is therefore not surprising 
that the window extends to larger values of $\Lambda_{\R}$.

By estimating the spectral density of the twisted mass Hermitean Dirac operator, the 
dimensionless quantity $r_0\Sigma^{1/3}$ was computed in Ref.~\cite{Cichy:2013gja}.
Since they have a smaller set of data, the analysis described in Section \ref{sec:seccont} 
is not a viable option for them. They opt for the strategy adopted in Ref.~\cite{Giusti:2008vb} 
which is inspired by NLO ChPT. They fit the mode number linearly in $M$ in the range 
$50$--$120$ MeV, and they extrapolate the results to the chiral and continuum limits 
linearly. The smaller quark masses and in particular the smaller values of 
$\Lambda_{\R}$ that we considered were instrumental to properly quantify and eventually 
reduce our systematic error. 

\subsection{Finite-size effects}
We estimate finite-volume effects using NLO ChPT (see Appendix \ref{app:modenumber:chpt}), 
and choose the parameters such that they are negligible 
within the statistical accuracy. For the lattice E5 we have explicitly checked 
that finite-size effects are within the expectations of ChPT by comparing the values 
of the mode number with those obtained on a lattice of $48\times 24^3$, lattice 
D5 in Table~\ref{tab:lambda2} of Appendix \ref{sec:parms}.

\section{Results and conclusions}
Our results show that in QCD with two flavours the low modes of the Dirac 
operator do condense in the continuum limit as expected by the 
Banks--Casher relation in presence of spontaneous symmetry breaking. The 
spectral density of the Dirac operator in the chiral limit at the origin 
is $[\pi\rho^\msbar(2\, \mbox{GeV})]^{1/3}= 261(6)(8)~\mbox{MeV}$, where the 
first error is statistical and the second is systematic. The latter is estimated 
so that the results from various fits are within the range covered by the 
systematic error: in particular the smaller value that we find in 
Section~\ref{sec:seccont} when a $\Lambda^2_R$ term is included in the fit function, 
and the higher one obtained in Section~\ref{sec:seccombined} when some of the data 
at the coarser lattice spacing are excluded from the fit. From the GMOR relation the 
best value of the chiral condensate that we obtain is 
$[\Sigma^\msbar_{\rm GMOR}(2\, \mbox{GeV})]^{1/3}= 263(3)(4)$~MeV, 
where again the first 
error is statistical and the second is systematic. The spectral density
at the origin thus agrees with $M_\pi^2 F_\pi^2/(2m_{\R})$ when both are extrapolated 
to the chiral limit. 

For the sake of clarity, the above values of the condensate have been expressed 
in physical units by 
supplementing the theory with a quenched ``strange'' quark, and by fixing 
the lattice spacing from the kaon decay constant $F_K$. They are therefore
affected by an intrinsic ambiguity due to the matching of $F_K$
in the $N_f=2$ partially quenched theory with its experimental value. The 
renormalisation group-invariant dimensionless ratio
\be
\frac{[\Sigma^{\rm RGI}]^{1/3}}{F} =  2.77(2)(4)\; ,
\ee
however, is a parameter-free prediction of the $N_f=2$ theory. It belongs to the family 
of unambiguous quantities that should be used for comparing computations 
in the two flavour theory rather than those expressed in physical units~\cite{Aoki:2013ldr}.

\acknowledgments{
Simulations have been performed on BlueGene/Q at CINECA 
(CINECA-INFN agreement), on HLRN, on JUROPA/JUQUEEN at 
J\"ulich JSC, on PAX at DESY--Zeuthen, and on Wilson at Milano--Bicocca. 
We thank these institutions for the computer resources and the technical 
support. We are grateful to our colleagues within the CLS initiative for 
sharing the ensembles of gauge configurations. G.P.E.~and L.G.~acknowledge 
partial support by the MIUR-PRIN contract 20093BMNNPR and by the INFN SUMA project. 
S.L.~and R.S.~acknowledge support by the DFG Sonderforschungsbereich/Transregio SFB/TR9.
}

\appendix

\section{Lattice action and operators}\label{app:A}
The gluons are discretized with the  Wilson plaquette 
action, while the doublet of mass-degenerate quarks
with the O$(a)$-improved Wilson action\footnote{The correction 
proportional to $b_{\rm g}$ is 
neglected.}~\cite{Sheikholeslami:1985ij,Luscher:1996sc} 
with its coefficient $c_\mathrm{sw}$ determined 
non-perturbatively~\cite{Jansen:1998mx}. We are interested in the flavour
non-singlet ($r,s=1,2$; $r\neq s$) fermion bilinears 
\be
P^{rs} = \psibar_{\; r} \gamma_5 \psi_s\;, \qquad 
A_0^{rs} = \psibar_r \gamma_0 \gamma_5 \psi_s\; .
\ee
The corresponding O$(a)$-improved renormalised operators are given by
\bea
P_{\rm R}^{rs} & = & Z_{\rm P}\, (1+ ({\bar b}_{\rm P} + {\tilde b}_{\rm P})\, a m)\,
P^{rs}\nonumber\; ,\\[0.125cm]
A_{0,\rm R}^{rs} & = & Z_{\rm A}\, (1+ ({\bar b}_{\rm A} + {\tilde b}_{\rm A})\, a m)\, 
\left\{A_0^{rs} + c_{\rm A}\, \frac{a}{2}\,(\partial_0^* + 
\partial_0)\, P^{rs} \right\}\; ,\label{eq:AR} 
\eea
where $\partial_0$ and $\partial_0^*$ are the forward and the backward 
lattice derivatives respectively. The coefficient $c_{\rm A}$ has been 
determined non-perturbatively for the $N_f=2$ theory in 
Ref.~\cite{DellaMorte:2005se}, while the $b$-coefficients are known in 
perturbation theory up to one loop only~\cite{Sint:1997jx,Sint:1997dj}.
The multiplicative renormalization constants $Z_{\rm A}$ and $Z_{\rm P}$ 
have been computed non-perturbatively in Ref.~\cite{Fritzsch:2012wq}. 
For the lattices considered in this paper, the numerical values of the 
improvement coefficients and of the renormalization constants are 
summarized in Table~\ref{tab:impr}. 
\begin{table}
\small
\begin{center}
\begin{tabular}{@{\extracolsep{0.0cm}}ccccccccccc}
\hline
$\beta$& run & $c_{\rm SW}$ & $c_{\rm A}$ & ${\tilde b}_{\rm P}$ & ${\tilde b}_{\rm A}$ & 
${\bar b}_\mu$& $Z_{\rm P}$ & $Z_{\rm A}$\\
\hline
5.2  &all&2.01715  & -0.06414 & 1.07224 & 1.07116 & -0.576 & 0.5184(53) &0.7703(57)\\[0.125cm]
5.3  &all&1.90952  & -0.05061 & 1.07088 & 1.06982 & -0.575 & 0.5184(53) &0.7784(52)\\[0.125cm]
5.5  &N5&1.751496  & -0.03613 & 1.06830 & 1.06728 & -0.572 & 0.5184(53) &0.7932(43)\\[0.125cm]
5.5  &N6,O7&1.751500  & -0.03613 & 1.06830 & 1.06728 & -0.572 & 0.5184(53) &0.7932(43)\\[0.125cm]
\hline
\end{tabular}
\end{center}
\caption{\label{tab:impr} Improvement coefficients and renormalization
constants for the $\beta$ values considered in the paper.}
\end{table}
The matching factors between $Z_{\rm P}$ in the Schr\"odinger functional
scheme and the renormalization-group invariant $Z^{\rm RGI}_{\rm P}$ 
(with the overall 
normalization convention of Ref.~\cite{Fritzsch:2012wq}) and
$Z^{\msbar}_{\rm P}(2~\mbox{GeV})$ are 
\be
Z^{\rm RGI}_{\rm P} = \frac{1}{1.308(16)}\, Z_{\rm P}\;, \qquad
Z^{\msbar}_{\rm P}(2~\mbox{GeV})= \frac{1}{0.740(12)}\, Z^{\rm RGI}_{\rm P}\; . 
\ee  
Using the PCAC relation, we can define
\be
m(x_0)=\frac{\frac{1}{2}(\partial_0 +\partial_0^*) 
  f_\mathrm{AP}(x_0)+ c_{\rm A} a
  \partial_0^* \partial_0 f_\mathrm{PP}(x_0)}{2 f_\mathrm{PP}(x_0)}\; ,
\label{eq:m}
\ee
where 
\bea
f_\mathrm{PP}(x_0)& = &-a^3\sum_{\vec{x}} 
\langle P^{12}(x) P^{21}(0)\rangle\; ,\nonumber\\[0.125cm]
f_\mathrm{AP}(x_0)& = & -a^3\sum_{\vec{x}} \langle A_0^{12}(x) P^{21}(0) 
\rangle\; . 
\label{eq:2pt}
\eea
At asymptotically large values of $x_0$, the mass $m(x_0)$ has a 
plateau which defines the value of $m$ to be used in Eqs.~(\ref{eq:AR}).
From this the renormalized quark mass is obtained as  
\be\label{eq:mR}
m_{\rm R}= \frac{Z_{\rm A}\, (1+ ({\bar b}_{\rm A} + {\tilde b}_{\rm A})
\, a m)}
{Z_{\rm P}\, (1+ ({\bar b}_{\rm P} + {\tilde b}_{\rm P})\, a m )}\, m\; . 
\ee
The bare pseudoscalar decay 
constant is given by~\cite{DelDebbio:2007pz} 
\be 
{\cal F}_{\pi} = 2 m \frac{G_{\pi}}{M^2_{\pi}}\; , 
\ee
where  $G_{\pi}$ is extracted from the behaviour of 
the correlator $f_\mathrm{PP}(x_0)$ at asymptotically large values of $x_0$
\be
f_\mathrm{PP}(x_0) = \frac{G^2_{\pi}}{M_{\pi}} e^{-M_{\pi} x_0}\; . 
\ee
Thanks to Eq.~(\ref{eq:AR}), the 
pseudoscalar decay constant is finally given by  
\be\label{eq:Fps}
F_{\pi} = Z_{\rm A}\, (1+ ({\bar b}_{\rm A} + {\tilde b}_{\rm A})\, a m)\; {\cal F}_{\pi}\; . 
\ee

\section{Quark masses, pion masses and decay constants\label{app:mmpif}}
\begin{figure}
\begin{minipage}{0.35\textwidth}
\includegraphics[width=7.0 cm,angle=0]{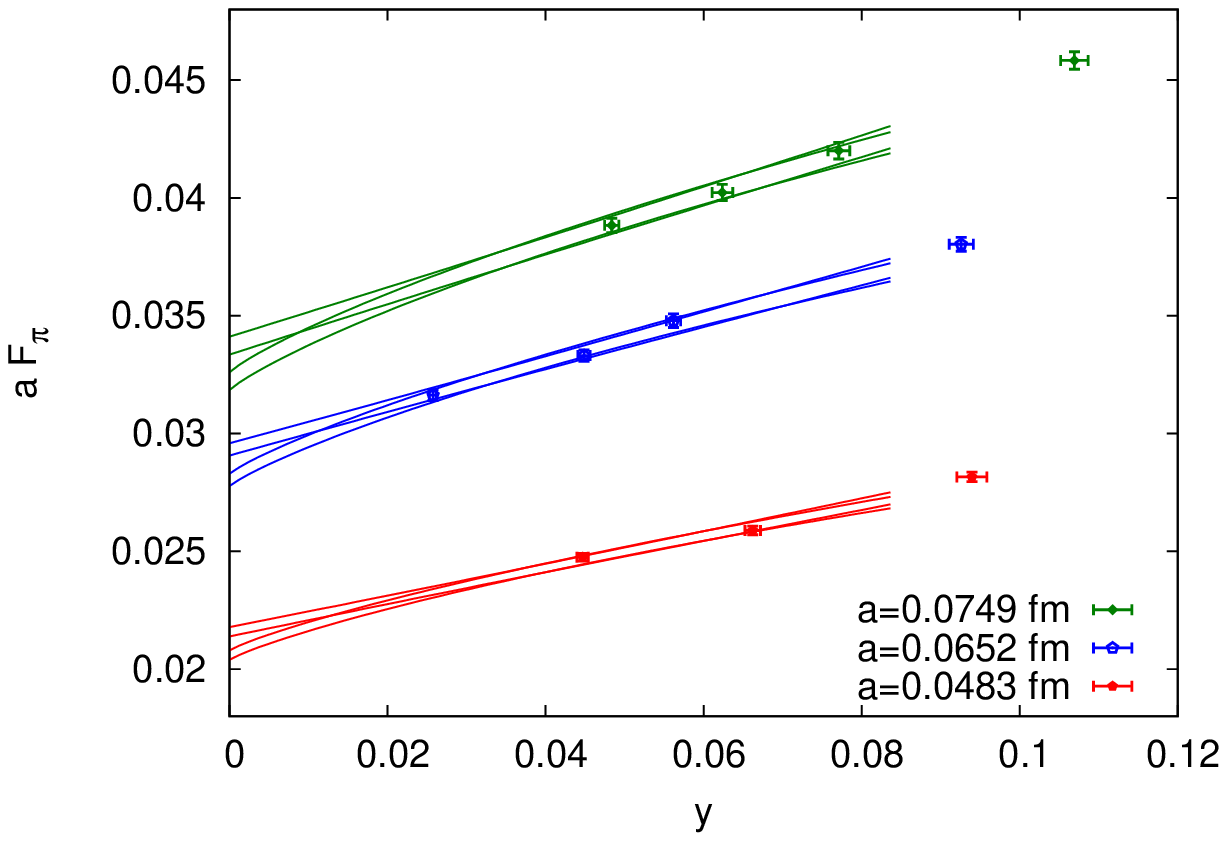}
\end{minipage}
\hspace{20mm}
\begin{minipage}{0.35\textwidth}
\includegraphics[width=7.0 cm,angle=0]{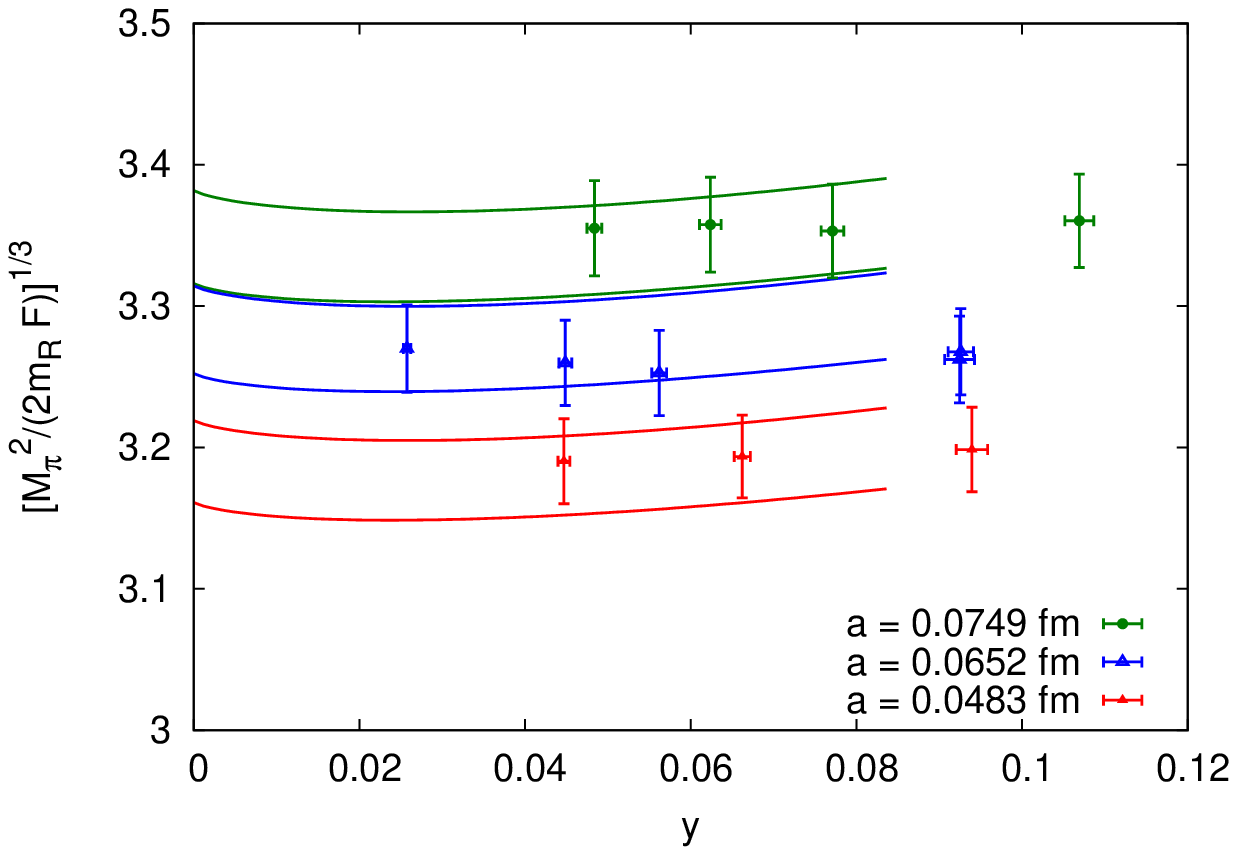}
\end{minipage}
\caption{Left: the pseudoscalar decay constant $a F_\pi$ versus 
\mbox{$y=M_\pi^2/(4 \pi F_\pi)^2$}. Right: The ratio 
$M_\pi^2/(2 m_R F)$ versus $y$. The bands are the result of a combined fit, 
see main text.}
\label{fig:FpiMpi}
\end{figure}

On all ensembles in Table~\ref{tab:ens} we have computed the two-point 
functions of the flavour non-singlet bilinears operators
in Eqs.~(\ref{eq:m}) and (\ref{eq:2pt}). They have been estimated
by using 10 to 20 $U(1)$ noise sources located on randomly chosen time 
slices. The bare quark mass $m(x_0)$ in Eq.~(\ref{eq:m}) has a plateau
for large enough $x_0$ over which we average. The pion mass $M_{\pi}$ and the bare 
pion decay constant ${\cal F}_{\pi}$
are extracted from $f_\mathrm{PP}(x_0)$ and the quark mass following 
Ref.~\cite{Fritzsch:2012wq}. In particular 
we determine the region $x_0 \in [x^{\rm min}_0; T - x^{\rm min}_0]$
where we can neglect the excited state contribution by first fitting
the pseudoscalar two-point function with a two-exponential fit
\be\label{eq:twoexp}
f_\mathrm{PP}(x_0) = d_1 \big[e^{-E_1 x_0} + e^{-E_1(T-x_0)}\big] + 
d_2\big[e^{-E_2 x_0} + e^{-E_2(T-x_0)}\big]
\ee
in a range where this function describes the data well for the given statistical 
accuracy. We then determine $x^{\rm min}_0$ to be the smallest value of $x_0$ where the 
statistical uncertainty on the effective mass
$m_{\rm eff}(x_0) = -\frac{{\rm d}}{{\rm d x_0}} \log[f_\mathrm{PP}(x_0)] $ 
is four times larger 
than the contribution of the excited state to $m_{\rm eff}(x_0)$ as given by the result of the fit.
In the second step only the first term of Eq.~(\ref{eq:twoexp}) is fitted to the 
data restricted to this region, and $E_1$ and $d_1$ are determined. The pion mass and its 
decay constant are then fixed to be 
$M_{\pi}=E_1$ and ${\cal F}_{\pi} = 2 \sqrt{d_1} m/M_{\pi}^{3/2}$ respectively.
\begin{table}
\small
\begin{center}
\begin{tabular}{@{\extracolsep{0.0cm}}clll}
\hline
id &~~~~$am$ &~~~$a M_{\pi}$ &~~~~$a F_{\pi}$ \\
\hline
A3 & 0.00985(6) & 0.1883(8)    & 0.04583(37) \\
A4 & 0.00601(6) & 0.1466(8)    & 0.04200(35) \\
A5 & 0.00444(6) & 0.1263(11)   & 0.04023(34)\\
B6 & 0.00321(4) & 0.1073(8)    & 0.03883(31)\\
\hline
E5 & 0.00727(3)   & 0.1454(5)  & 0.03803(29) \\
F6 & 0.00374(3)   & 0.1036(5)  & 0.03479(29) \\
F7 & 0.002721(20) & 0.0886(4)  & 0.03331(24) \\
G8 & 0.001395(18) & 0.0638(4)  & 0.03162(23) \\
   \hline
N5 & 0.00576(3)   & 0.1085(8)  & 0.02816(21) \\
N6 & 0.003444(15) & 0.0837(3)  & 0.02589(19) \\
O7 & 0.002131(9) & 0.06574(23) & 0.02475(16) \\
\hline
\end{tabular}
\end{center}
\caption{\label{tab:spect} The bare quark mass $a m$ as defined in 
Eq.~(\ref{eq:m}), the pion mass $a M_{\pi}$ and pion decay
constant $a F_{\pi}$ as defined in Eq.~(\ref{eq:Fps}). }
\end{table}
The numerical results for all lattices are reported in 
Table~\ref{tab:spect}, and those for the pseudoscalar decay 
constant and for the cubic root of the ratio $M^2_{\pi}/(2m_R F)$ 
are shown in Fig.~\ref{fig:FpiMpi} versus $y=M_{\pi}^2/(4 \pi F_{\pi})^2$. We fit 
$F_{\pi}$ to the function 
\be
a F_{\pi} = (a F)\, \{1 - y \ln(y) + b y\}\; ,
\ee 
where $b$ is common to all lattice spacings, restricted to the points 
with $M_{\pi}<\, 400$~MeV (see left plot of Fig.~\ref{fig:FpiMpi}). This
function rests on the Symanzik expansion and is compatible 
with Wilson ChPT (WChPT) at the NLO \cite{Aoki:2009ri}. To estimate 
the systematic error, we performed a number of fits to different functions: 
linear in $y$ with $M_{\pi}<\, 400$~MeV, and next-to-next-to-leading
order in ChPT with all data included. As a final result we quote 
$a F=0.0330(4)(8)$, $0.0287(3)(7)$ and $0.0211(2)(5)$ at $a=0.075$, $0.065$ and $0.048$~fm 
respectively, where the second (systematic) error takes into account the spread of the 
results from the various fits. By fixing the scale from $F_K$, and by 
performing a continuum-limit extrapolation we obtain our final result $F=85.8(7)(20)$~MeV.

We further compute the ratio $M_{\pi}^2/(2m_R F)$ 
for all data points. We fit the data restricted to $M_{\pi}<\, 400$~MeV to 
\be\label{eq:Mpicont}
\Big[\frac{M_{\pi}^2}{2m_R F}\Big]^{1/3}=(s_0+s_1 (aF)^2)\{1+ \frac{y}{6} \ln(y) +d\, y\}\; , 
\ee
where $s_0$, $s_1$ and $d$ are common to all lattice spacings,
and the fit function is again the one resting on the Symanzik expansion 
and  compatible with WChPT at the NLO. Also in this case we checked several 
variants although the data look very flat up to the heaviest mass.
From the fits we get $s_0=3.06(3)(4)$, where the systematic error is determined as for $F$. 
This translates to a value for the renormalisation-group-invariant dimensionless ratio 
of $[\Sigma^{\rm RGI}]^{1/3}/F =2.77(2)(4)$, which in turn corresponds to 
$[\Sigma^\msbar(2\, \mbox{GeV})]^{1/3} =263(3)(4)$~MeV if again $F_K$ is used to 
set the scale. 

\section{Mode number in chiral perturbation theory }\label{app:modenumber:chpt}
When chiral symmetry is spontaneously broken, the mode number
can be computed in the chiral effective theory. At the NLO it reads~\cite{Giusti:2008vb} 
(see also Ref.~\cite{Giusti:2008paa})
\be\label{eq:RNLO}
\nu^{\rm nlo}(\Lambda_{\rm R},m_{\rm R}) =  \frac{2 \Sigma \Lambda_{\rm R} V}{\pi}  \Big\{ 1 + 
\frac{m_{\rm R} \Sigma}{(4\pi)^2 F^4}\Big[3\, \bar l_6 + 1 - \ln(2) 
- 3 \ln\Big(\frac{\Sigma m_{\rm R}}{F^2 \bar\mu^2}\Big) + 
f_\nu\left(\frac{\Lambda_{\rm R}}{m_{\rm R}}\right)\Big]\Big\}\; , 
\ee
where 
\be
f_\nu(x) = x \left[{\rm arctan}(x) - \frac{\pi}{2}\right]
- \frac{1}{x} {\rm arctan}(x) - \ln(x) - \ln(1+x^2)\; .
\ee
The constants $F$ and $\bar l_6$ are, respectively, the pion decay
constant in the chiral limit and a SU$(3|1)$ low-energy effective coupling
renormalized  at the scale $\bar\mu$. The formula in Eq.~(\ref{eq:RNLO}) has 
some interesting properties:
\begin{itemize}
\item for $x \rightarrow \infty$
\be
f_\nu(x) \longrightarrow_{\!\!\!\!\!\!\!\!\!\!\!\!_{x\rightarrow\infty}} -3 \ln(x)\;,
\ee
and therefore at fixed $\Lambda_{\rm R}$ the mode number has no chiral 
logs when $m_{\rm R}\rightarrow 0$;
\item since in the continuum the operator $D^\dagger_m D_m$ has a threshold 
at $\alpha=m^2$, the mode number must satisfy  
\be
\lim_{\Lambda_{\rm R}\rightarrow 0} \nu^{\rm nlo}(\Lambda_{\rm R},m_{\rm R}) = 0\; ,
\ee
a property which is inherited by the NLO ChPT formula;
\item in the chiral limit $\nu^{\rm nlo}(\Lambda_{\rm R},m_{\rm R})/\Lambda_{\rm R}$ 
{\it becomes independent on} $\Lambda_{\rm R}$. This is an accident of the 
$N_f=2$ ChPT theory at NLO \cite{Smilga:1993in};
\item the $\Lambda_{\rm R}$-dependence in the square brackets on the r.h.s. 
of (\ref{eq:RNLO}) is parameter-free. Since 
$\frac{m_{\rm R} \Sigma^2}{(4\pi)^2 F^4}>0$, the behaviour of the 
function $f_\nu(x)$ implies that $\nu^{\rm nlo}(\Lambda_{\rm R},m_{\rm R})/\Lambda_{\rm R}$ is a 
decreasing function of $\Lambda_{\rm R}$ at fixed $m_{\rm R}$, and no ambiguity is 
left due to free parameters. 
\end{itemize}
\vspace{0.125cm}

\noindent At the NLO the effective spectral density defined in 
Eq.~(\ref{eq:discD}) reads
\be\label{eq:sigmatilde}
\wS^{\rm nlo} =  \Sigma \Big\{ 1 + \frac{m_{\rm R} \Sigma}{(4\pi)^2 F^4}
\Big[3\, \bar l_6 + 1 - 
\ln(2) - 3 \ln\Big(\frac{\Sigma m_{\rm R}}{F^2 \bar\mu^2}\Big) + 
\tilde g_\nu\left(\frac{\Lambda_{1,\R}}{m_{\R}},\frac{\Lambda_{2,\R}}{m_{\R}}\right)
\Big]\Big\}\, , 
\ee
where 
\be
\tilde g_\nu\left(x_1,x_2\right) = \frac{f_\nu(x_1)+f_\nu(x_2)}{2} + 
\frac{1}{2}\,\frac{x_1+x_2}{x_2-x_1}\,\Big[f_\nu(x_2)-f_\nu(x_1)\Big]\; .
\ee
The quantity ${\wS}^{\rm nlo}$ 
inherits the same peculiar properties 
of $\nu^{\rm nlo}(\Lambda_{\R},m_{\R})/\Lambda_{\R}$ at NLO: at fixed 
$\Lambda_{1,\R}$ and $\Lambda_{2,\R}$ it has no chiral logarithms when 
$m_{\R}\rightarrow 0$, it is independent from $\Lambda_{1,\R}$ and 
$\Lambda_{2,\R}$ in the 
chiral limit, and at non-zero quark mass it is a decreasing 
parameter-free (apart the overall factor) function of 
$(\Lambda_{1,\R} + \Lambda_{2,\R})/2$. It is very weakly dependent 
on $(\Lambda_{1,\R}-\Lambda_{2,\R})$ in the range we are interested 
in. To have a quantitative idea of the 
$(\Lambda_{1,\R} + \Lambda_{2,\R})/2$ dependence of $\wS^{\rm nlo}$
we can choose $\Sigma=(260~\mbox{MeV})^3$, $F=85$~MeV, 
$m^{\rm sea}_{\R}=10$~MeV, $\Lambda_{1,\R}=20,40$~MeV, 
$\Lambda_{2,\R}=25,55$~MeV to obtain
\be
\frac{\Sigma}{(4\pi)^2 F^4} = 0.00213~\mbox{MeV}^{-1}\; , \quad
0.0213\cdot\left[
\tilde g_\nu\Big(\frac{20}{10},\frac{25}{10}\Big) -
\tilde g_\nu\Big(\frac{40}{10},\frac{55}{10}\Big)\right] = 0.0467\, .
\ee
For light values of the quark masses the variations are rather mild
, i.e. of the order of few percent.
The next-to-next-to leading corrections in $\wS$ are 
of the form ${\cal O}(\Lambda_{\R}^2, m_{\R} \Lambda_{\R},m_{\R}^2)$. They are expected 
to spoil some of the peculiar properties of the NLO formula.
In the chiral limit the ${\cal O}(\Lambda_{\R}^2)$ corrections can 
induce a $\Lambda_{\R}$-dependence, and the 
${\cal O}(m_{\R} \Lambda_{\R})$ can change the parameter-free dependence 
on $\Lambda_{\R}$ within the square brackets on the r.h.s.~of 
Eq.~(\ref{eq:sigmatilde}). 

\subsection{Finite volume effects}\label{sec:finvol}
Finite volume effects in the mode number were computed in the chiral 
effective theory at the NLO in Refs.~\cite{Giusti:2008vb,Giusti:2008paa}
(see also \cite{Necco:2011vx}). They are given by
\bea
\left(\frac{\Delta \nu_{V}}{\nu}\right)^{\rm{nlo}} & = & \frac{\Sigma}{(4\pi)^2 F^4}
\sum_{\{n_1,\dots,n_4\}}\,\!\!\!\!\!\!\!\! '\, \lim_{\epsilon\rightarrow 0}
\left\{\frac{2}{\Lambda_{\rm R}} {\rm Im} 
\Big[F_{-2}\left(\frac{\Sigma q_n^2}{4F^2},i\Lambda_{\rm R}+m_{\rm R}+\epsilon\right)\Big] - 
\right. \nonumber \\[0.25cm]
&& \left. \frac{m_{\rm R}}{\Lambda_{\rm R}}{\rm Im}\Big[F_{-1}
\Big(\frac{\Sigma q_n^2}{2F^2},i\Lambda_{\rm R}+\epsilon\Big)\Big]+ 
{\rm Re}\Big[F_{-1}\Big(\frac{\Sigma q_n^2}{2F^2},i\Lambda_{\rm R}+\epsilon\Big)\Big]\; 
 \right\}\;, 
 \label{eq:chpt:finvol}
\eea
where
\be\displaystyle\label{eq:Fnu}
F_\nu(b,z) = 2\, \left(\frac{b}{z}\right)^{\nu/2}\, K_\nu(2 \sqrt{b z})\; ,
\ee
with ${\rm Re}\, b>0$, ${\rm Re}\, z>0$, and $K_\nu$ is a 
modified Bessel function \cite{Abramowitz:1970}. 
Furthermore, $q_n^2 = \sum_{\mu=1}^{d} (n_\mu L_\mu)^2$ and
$\sum_{\{n_1,\dots,n_d\}}'$ denotes the sum over all integers without $n=(0,\dots,0)$.
By expanding the Bessel functions for large arguments~\cite{Abramowitz:1970},
it is straightforward to show that the most significant terms in the sum 
on the r.h.s of Eq.~(\ref{eq:chpt:finvol}) are proportional 
to the exponentials $\exp\{-M_1 L/\sqrt{2}\}$ and $\exp\{- M_{2} L/2\}$,  
where $M_1$ and $M_2$ are the leading-order expressions in 
ChPT for the mass of a pseudoscalar meson made of two valence quarks of mass 
$\Lambda_{\R}$ and $(\sqrt{\Lambda_{\R}^2+m_{\R}^2}+m_{\R})$ respectively.

\subsection{Discretization effects\label{eq:WChPT}}
At finite lattice spacing and volume, the threshold region should 
be treated carefully in ChPT~\cite{Damgaard:2010cz}. 
The latter can be avoided by considering the 
quantity $\wS$, 
with $\Lambda_{2,\R}>\Lambda_{1,\R}\gg 1/\Sigma V$. In this 
case the computation in the GSM power-counting regime of the 
Wilson ChPT gives~\cite{Necco:2011vx}  
\be\label{eq:sigmatildelat}
\wS^{\rm nlo}(a) = \wS^{\rm nlo} 
-32 \,(W_0 a)^2\, W_8' m_{\rm R} 
\frac{1}{\Lambda_{1,\R} \Lambda_{2,\R}}\; . 
\ee
Since $W_8'$ is expected to be negative \cite{Hansen:2011kk,Splittorff:2012gp}, if we rewrite 
\be
\Lambda_{1,\R} \Lambda_{2,\R} =\left(\frac{\Lambda_{1,\R}
+\Lambda_{2,\R}}{2}\right)^2
-\frac{1}{4}(\Lambda_{2,\R}-\Lambda_{1,\R})^2\; 
\ee
and we keep constant $(\Lambda_{2,\R}-\Lambda_{1,\R})$, then 
$\wS^{\rm nlo}(a)$ is a decreasing 
function of $\Lambda_{\R}=(\Lambda_{2,\R}+\Lambda_{1,\R})/2$ on the lattice too. 
At variance with the continuum case, however, a free parameter 
$W_0^2 W_8'$ appears
in the function, and its magnitude cannot be predicted. Remarkably 
$\wS^{\rm nlo}(a)$ is free from discretization effects
in the chiral limit, and therefore it is independent on 
$\Lambda_{1,\R}$ and $\Lambda_{2,\R}$.
The continuum extrapolation of the chiral value 
of $\wS^{\rm nlo}(a)$ then removes the discretization 
effects due to the reference scale used.

\section{Numerical results for the mode number}\label{sec:parms}
We collect the results for the mode number in Tables \ref{tab:lambda},
\ref{tab:lambda2} and  \ref{tab:lambda3}. For each lattice 
the values of $a M$ correspond to approximatively 
$\Lambda_{\rm R}=$20, 25, 30, 40, 55, 71, 86, 101, 116 MeV with the exception 
of the lattice E5 for which also $\Lambda_{\rm R}=151, 202, 303, 505$~MeV 
were computed. 

\begin{table}[hb]
\small
\begin{center}
\begin{tabular}{@{\extracolsep{0.4cm}}cccccccc}
\hline
id & $N_{\text{cnfgs}}$ & $aM$         & $\nu$    \\ 
\hline
A3  & 55        & 0.008673     & 13.3(6)                    \\ 
    &           & 0.009208     & 16.2(6)                    \\ 
    &           & 0.009821     & 20.5(7)                   \\ 
    &           & 0.011235     & 29.6(9)                    \\ 
    &           & 0.013665     & 47.3(10)                    \\ 
    &           & 0.016322     & 66.9(12)                    \\ 
    &           & 0.019110     & 88.2(14)                    \\ 
    &           & 0.021979     & 111.1(16)                    \\ 
    &           & 0.024901     & 134.6(18)                    \\ 
A4  & 55        & 0.006205     & 11.6(6)                    \\ 
    &           & 0.006929     & 15.9(7)                    \\ 
    &           & 0.007723     & 20.6(7)                    \\ 
    &           & 0.009447     & 30.8(8)                    \\ 
    &           & 0.012228     & 48.8(10)                    \\ 
    &           & 0.015127     & 68.6(12)                    \\ 
    &           & 0.018088     & 89.6(13)                    \\ 
    &           & 0.021085     & 110.9(15)                    \\ 
    &           & 0.024103     & 132.5(15)                    \\ 
A5  & 55        & 0.005352     & 11.4(6)                     \\ 
    &           & 0.006176     & 15.6(6)          \\ 
    &           & 0.007054     & 20.6(7)          \\ 
    &           & 0.008905     & 31.9(8)          \\ 
    &           & 0.011810     & 50.1(11)         \\ 
    &           & 0.014786     & 68.3(13)         \\ 
    &           & 0.017799     & 88.7(14)         \\ 
    &           & 0.020831     & 108.7(16)         \\ 
    &           & 0.023877     & 129.2(18)         \\ 
B6  & 50        & 0.004800     & 59.5(10)                    \\ 
    &           & 0.005703     & 82.5(11)                    \\ 
    &           & 0.006642     & 108.4(13)                    \\ 
    &           & 0.008580     & 162.3(16)                    \\ 
    &           & 0.011563     & 253.0(22)                    \\ 
    &           & 0.014586     & 346.5(25)                    \\ 
    &           & 0.017629     & 443(3)                    \\ 
    &           & 0.020683     & 543(3)                    \\ 
    &           & 0.023743     & 647(4)                    \\ 
\hline
\end{tabular}
\end{center}
\caption{\label{tab:lambda} 
Values of $aM$ and the corresponding results for $\nu$ 
for each lattice at $\beta=5.2$.}
\end{table}

\begin{table}
\small
\begin{center}
\begin{tabular}{@{\extracolsep{0.4cm}}ccccccc}
\hline
id & $N_{\text{cnfgs}}$ & $aM$         & $\nu$    \\ 
\hline
D5  & 345        & 0.006720 &2.09(9) \\ 
    &           & 0.007239     &2.77(10) \\ 
    &           & 0.007826     &3.42(10) \\ 
    &           & 0.009153     &5.26(12) \\ 
    &           & 0.011385     &8.38(16) \\ 
    &           & 0.013782     &11.69(19) \\ 
    &           & 0.016271     &15.16(22) \\ 
    &           & 0.018815     &18.61(25) \\ 
    &           & 0.021396     &22.3(3) \\ 
E5  & 92        & 0.006720     & 7.3(3)                   \\ 
    &           & 0.007239     & 9.3(3)                    \\ 
    &           & 0.007826     & 11.5(3)                    \\ 
    &           & 0.009153     & 17.1(4)                    \\ 
    &           & 0.011385     & 26.9(5)                    \\ 
    &           & 0.013782     & 37.4(7)                    \\ 
    &           & 0.016271     & 47.3(8)                    \\ 
    &           & 0.018815     & 58.0(9)                    \\ 
    &           & 0.021396     & 68.8(10)                    \\ 
    &           & 0.027499     & 93.7(10)                    \\ 
    &           & 0.036321     & 138.6(12)                    \\ 
    &           & 0.054110     & 259.7(16)                    \\ 
    &           & 0.089863     & 689(3)                    \\ 
F6  & 50        & 0.004618     &34.7(9)                     \\ 
    &           & 0.005342     &47.6(11)                     \\ 
    &           & 0.006111     &60.7(12)                    \\ 
    &           & 0.007732     &90.8(16)                     \\ 
    &           & 0.010268     &135.8(17)                     \\ 
    &           & 0.012865     &183.0(20)                     \\ 
    &           & 0.015492     &230.9(23)                     \\ 
    &           & 0.018137     &280(3)                     \\ 
    &           & 0.020791     &330(3)                     \\ 
F7  & 50        & 0.004159     & 34.7(9)                    \\ 
    &           & 0.004950     & 47.0(10)                    \\ 
    &           & 0.005770     & 59.3(10)                    \\ 
    &           & 0.007464     & 87.1(12)                    \\ 
    &           & 0.010065     & 128.9(16)                    \\ 
    &           & 0.012701     & 172.0(21)                    \\ 
    &           & 0.015354     & 217.2(23)                    \\ 
    &           & 0.018015     & 265(3)                    \\ 
    &           & 0.020682     & 314(3)                    \\ 
G8  & 50        & 0.003737     &113.7(16)                  \\ 
    &           & 0.004599     &153.8(18)                   \\ 
    &           & 0.005472     &196.7(22)                   \\ 
    &           & 0.007233     &282.3(25)                   \\ 
    &           & 0.009892     &409(3)                    \\ 
    &           & 0.012560     &543(3)                     \\ 
    &           & 0.015233     &682(4)                     \\ 
    &           & 0.017910     &828(4)                     \\ 
    &           & 0.020587     &981(5)                     \\ 
\hline
\end{tabular}
\end{center}
\caption{\label{tab:lambda2} 
As in Table \ref{tab:lambda} but for $\beta=5.3$.
}
\end{table}

\begin{table}
\small
\begin{center}
\begin{tabular}{@{\extracolsep{0.4cm}}ccccccc}
\hline
id & $N_{\text{cnfgs}}$ & $aM$         & $\nu$    \\ 
\hline
N5  & 60        & 0.005287     &12.0(6)                  \\ 
    &           & 0.005647     &15.6(6)                 \\ 
    &           & 0.006058     &19.3(7)                  \\ 
    &           & 0.006998     &27.3(8)                  \\ 
    &           & 0.008599     &40.2(9)                  \\ 
    &           & 0.010334     &52.3(10)                  \\ 
    &           & 0.012146     &65.0(11)                  \\ 
    &           & 0.014005     &77.7(12)                  \\ 
    &           & 0.015895     &91.2(13)                  \\ 
N6  & 60        & 0.003797     & 11.0(4)                    \\ 
    &           & 0.004284     & 14.9(5)                    \\ 
    &           & 0.004812     & 18.3(5)                    \\ 
    &           & 0.005949     & 25.6(7)                    \\ 
    &           & 0.007765     & 37.3(8)                    \\ 
    &           & 0.009646     & 49.1(8)                    \\ 
    &           & 0.011562     & 60.4(9)                    \\ 
    &           & 0.013496     & 72.6(10)                    \\ 
    &           & 0.015444     & 85.8(11)                    \\ 
O7  & 50        & 0.003137     & 34.3(9)                       \\ 
    &           & 0.003710     & 45.9(10)                 \\ 
    &           & 0.004309     & 57.5(11)                 \\ 
    &           & 0.005548     & 78.5(12)                 \\ 
    &           & 0.007459     & 111.9(15)                \\ 
    &           & 0.009399     & 147.8(16)               \\ 
    &           & 0.011354     & 184.0(18)               \\ 
    &           & 0.013316     & 220.8(19)               \\ 
    &           & 0.015284     & 260.2(21)               \\ 
\hline
\end{tabular}
\end{center}
\caption{\label{tab:lambda3} 
As in Table \ref{tab:lambda} but for $\beta=5.5$.
}
\end{table}

\begin{table}
\small
\begin{center}
\begin{tabular}{@{\extracolsep{0.4cm}}c|ccc}
\hline
$\Lambda_{\rm R}/m_{\rm R}$            &   12.9 	        & 20.9		& 32.0 \\
\hline
22.7					& 0.0289(20)		& 0.032(3)			& 0.033(3)		\\
27.7					& 0.0249(21)		& 0.023(3)			& 0.029(3)		\\
35.3					& 0.0191(16)		& 0.025(3)			& 0.0308(24)		\\
47.9					& 0.0192(15)		& 0.0239(22)		& 0.0288(19)		\\
63.0					& 0.0221(15)		& 0.0228(24)		& 0.0229(18)		\\
78.2					& 0.0210(16)		& 0.0174(20)		& 0.0224(18)		\\
93.3					& 0.0212(14)		& 0.0221(21)		& 0.0211(18)		\\
108.4					& 0.0237(15)		& 0.0257(22)		& 0.0243(19)		\\
\hline
\end{tabular}
\end{center}
\caption{\label{tab:sigmatildecont} 
The effective density $\wS$ in the continuum is given for various values of the cutoff $\Lambda_{\rm R}$ and the quark mass $m_{\rm R}$. 
These data are obtained by first interpolating $\wS$ linearly in $m_{\rm R}$ for each $\Lambda_{\rm R}$ and lattice spacing $a$, followed by an extrapolation linear in $a^2$ to the continuum for each pair of $(\Lambda_{\rm R},m_{\rm R})$, as described in Sections \ref{sec:firstlook} and \ref{sec:seccont}. 
$\wS$ is given in GeV$^3$, $\Lambda_{\rm R}$ and $m_{\rm R}$ are given in MeV. 
}
\end{table}

\clearpage

\section{Numerical analysis of discretization effects\label{app:disc}}
In this appendix we report more details on the discretization effects that we have observed 
in our data. We limit ourselves to an empirical discussion of the results obtained by 
following the strategy described in Section~\ref{sec:seccont}. 
\begin{figure}
\begin{minipage}{0.35\textwidth}
\includegraphics[width=7.0 cm,angle=0]{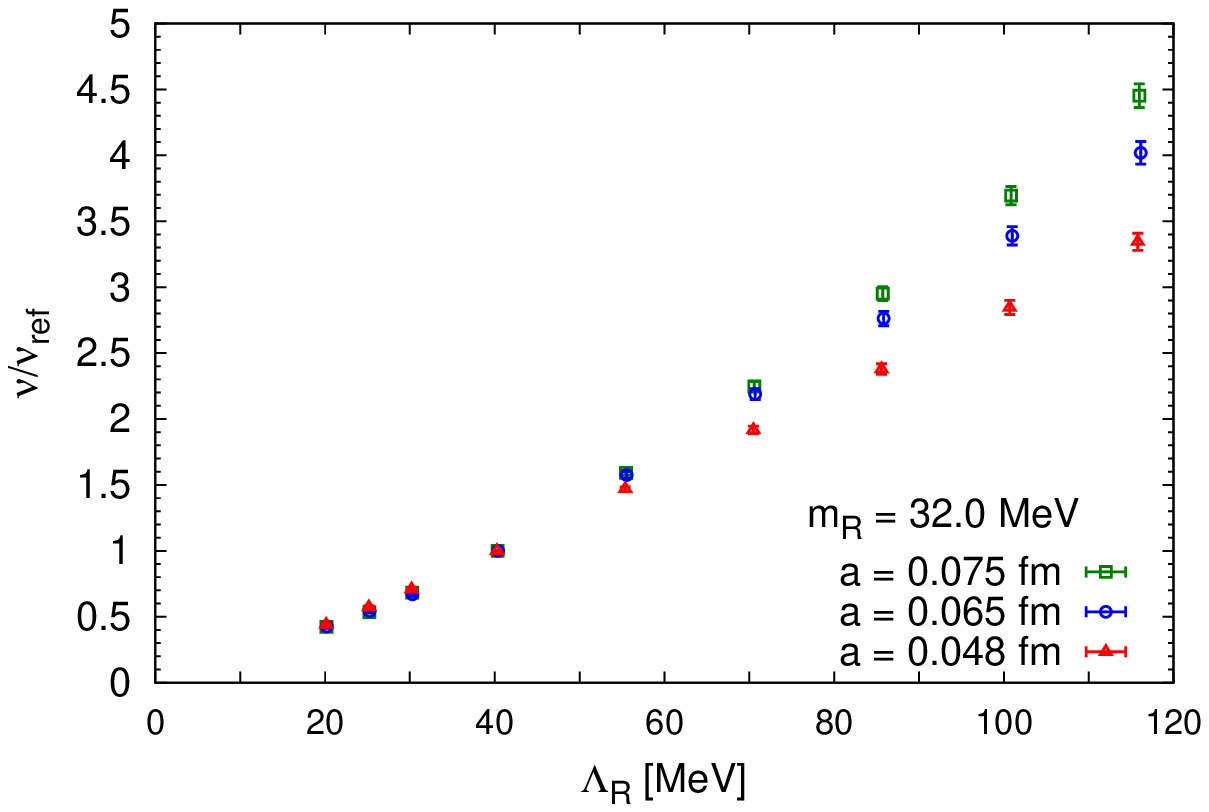}
\end{minipage}
\hspace{20mm}
\begin{minipage}{0.35\textwidth}
\includegraphics[width=7.0 cm,angle=0]{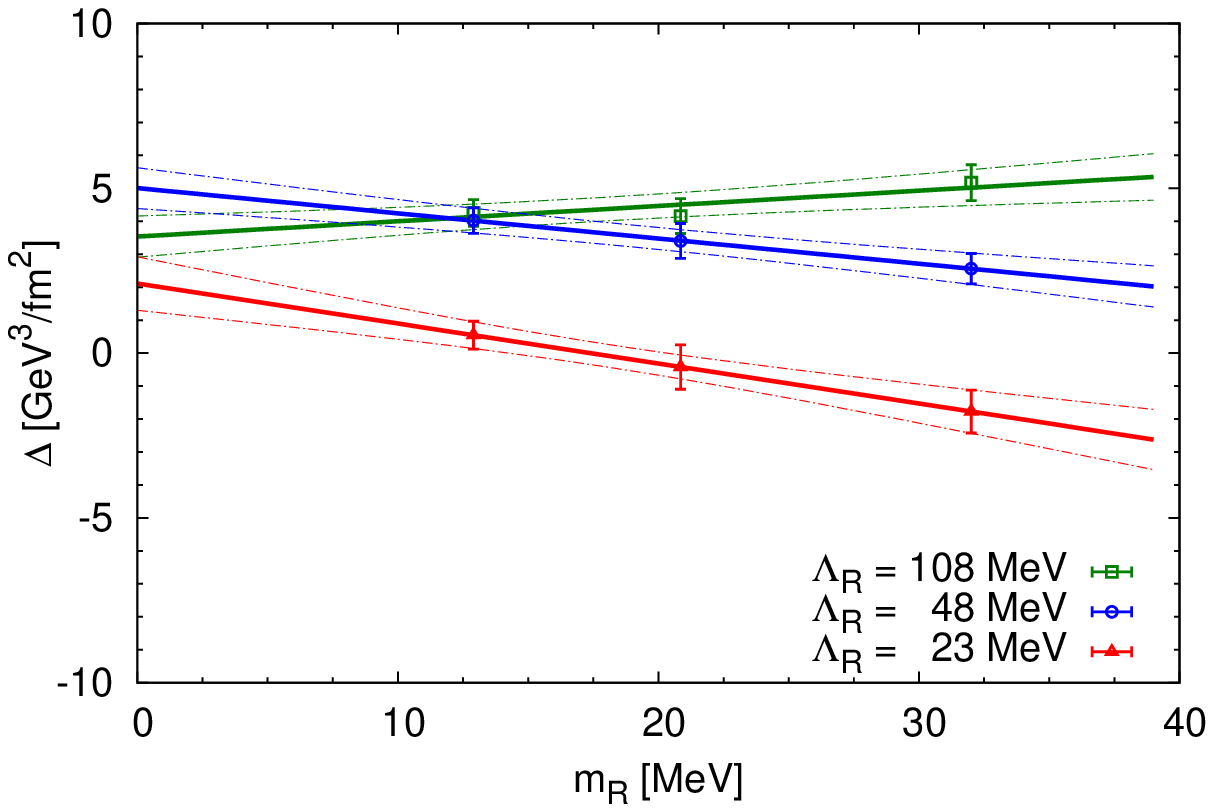}
\end{minipage}
\caption{Left: mode number at $m_{\rm R}=32$ MeV for all three lattice spacings and all cutoffs $\Lambda_{\rm R}$, normalized with respect to its value at $\Lambda_{\rm R}=40$ MeV. Right: discretization effects 
$\Delta$ of the effective spectral density as defined in Eq.~\eqref{eq:disc1}, shown 
vs.~$m_{\rm R}$ for three values of $\Lambda_{\rm R}$. The fit in the 
plot follows Eq.~\eqref{eq:disc2}, the resulting parameters of which are shown in 
Figure~\ref{fig:disc2}.}
\label{fig:disc1}
\end{figure}

\begin{figure}
\begin{minipage}{0.35\textwidth}
\includegraphics[width=7.0 cm,angle=0]{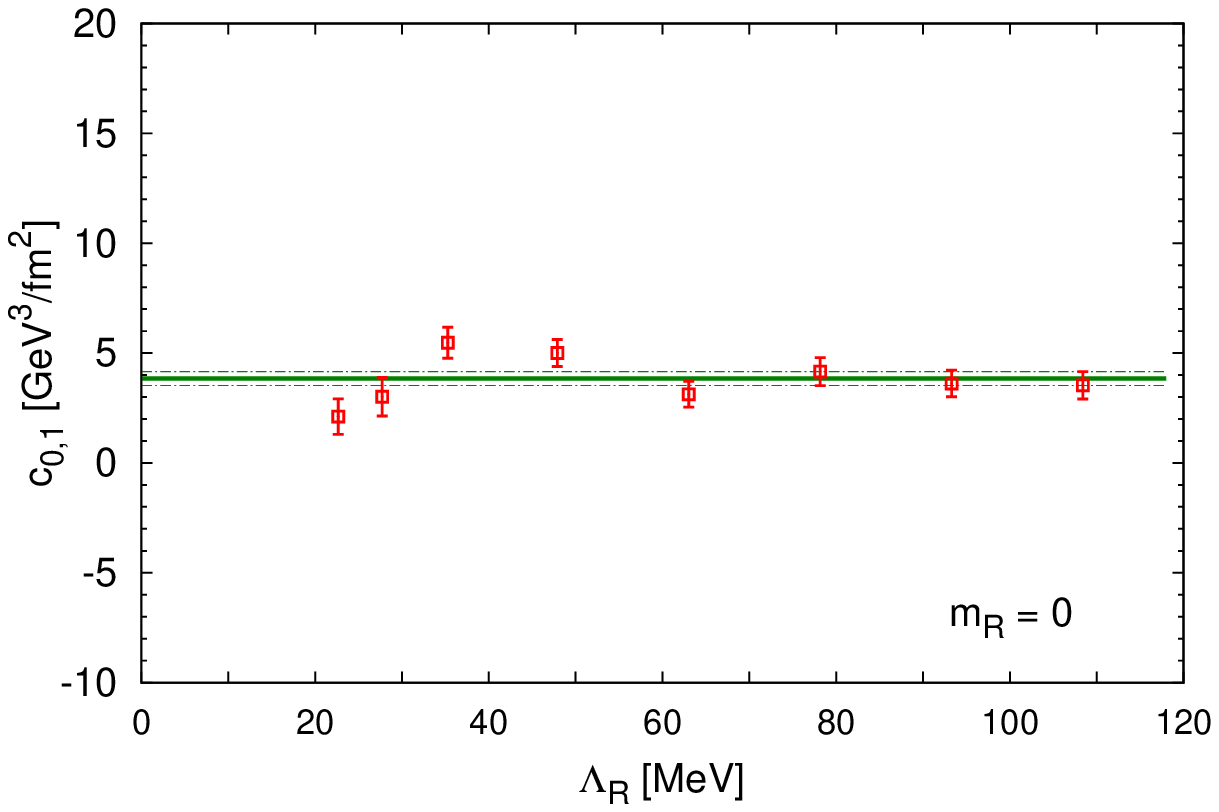}
\end{minipage}
\hspace{20mm}
\begin{minipage}{0.35\textwidth}
\includegraphics[width=7.0 cm,angle=0]{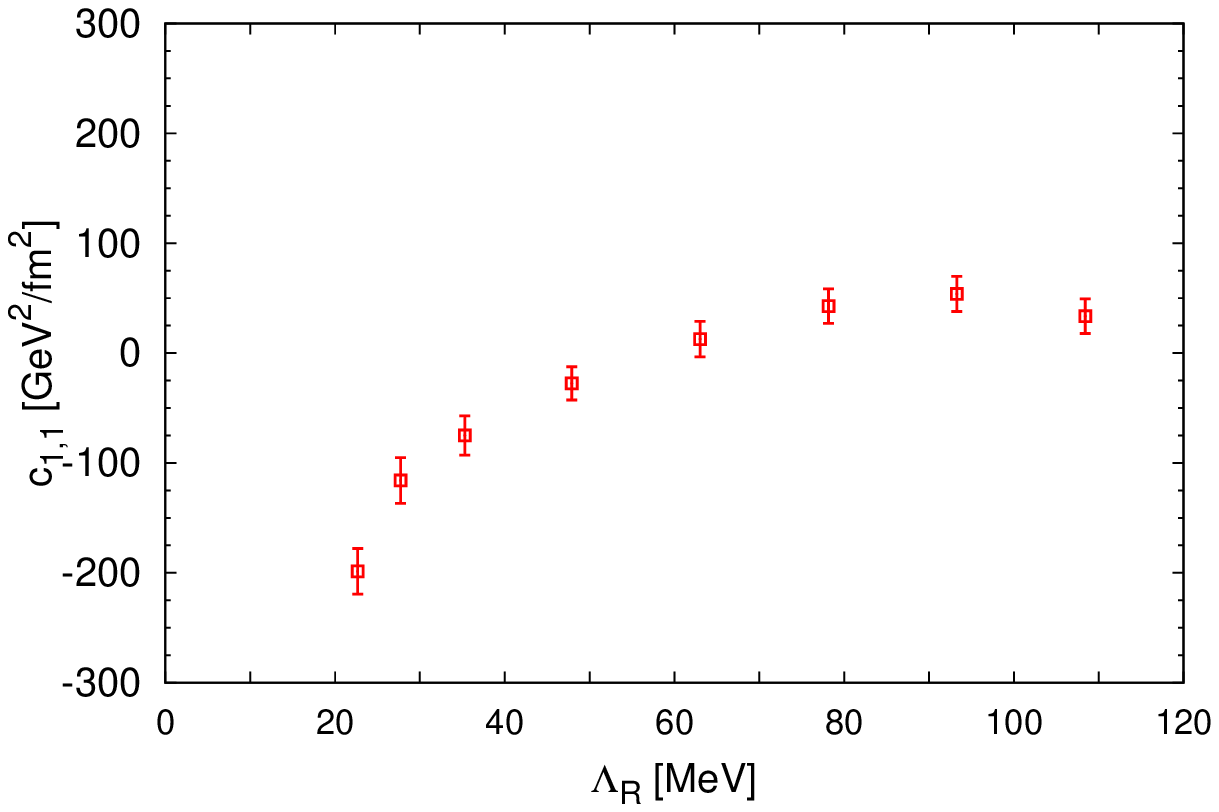}
\end{minipage}
\caption{Left: mass-independent discretization effects $c_{0,1}$ as defined in 
Eq.~\eqref{eq:disc2} vs.~$\Lambda_{\rm R}$. A fit of the plateau gives 3.8(3) GeV$^3$/fm$^2$.
Right: mass-dependent discretization effects $c_{1,1}$ as defined
in Eq.~\eqref{eq:disc2} (but with $c_{0,1}(\Lambda_{\rm R})$ constrained to be constant), as a function of $\Lambda_{\rm R}$.}
\label{fig:disc2}
\end{figure}

A first look into the data reveals that discretization effects in $\nu$ show a non-trivial 
dependence on $\Lambda_{\rm R}$ and $m_{\rm R}$. We plot the mode number at $m_{\rm R}=32$ MeV, 
normalized with respect to its value at $\Lambda_{\rm R}=40$ MeV, for all three lattice spacings 
and all values of $\Lambda_{\rm R}$ in Figure~\ref{fig:disc1}, left hand side. 
After interpolating the effective spectral density in $m_{\rm R}$, we fit the data linearly 
in $a^2$  
\be\label{eq:disc1}
\wS(\Lambda_{\rm R},m_{\rm R},a) = \wS(\Lambda_{\rm R},m_{\rm R},0) + a^2\Delta(\Lambda_{\rm R},m_{\rm R})
\ee 
for each pair of $(\Lambda_{\rm R},m_{\rm R})$. By fitting $\Delta$ linearly in $m_{\rm R}$
(Figure~\ref{fig:disc1}, right plot)
\be\label{eq:disc2}
\Delta(\Lambda_{\rm R},m_{\rm R}) = c_{0,1}(\Lambda_{\rm R}) + c_{1,1}(\Lambda_{\rm R})m_{\rm R}  
\ee
for each $\Lambda_{\rm R}$, we obtain the values for $c_{0,1}(\Lambda_{\rm R})$ shown in the left 
plot of Figure~\ref{fig:disc2}. Within errors, $c_{0,1}(\Lambda_{\rm R})$ turns out to be 
compatible with a constant. To reduce the noise in $c_{1,1}(\Lambda_{\rm R})$, we repeat the fit
in Eq.~(\ref{eq:disc2}) but constraining $c_{0,1}(\Lambda_{\rm R})$ to be a constant.  
The results of this fit are shown in the right plot of Figure~\ref{fig:disc2}. The coefficient 
$c_{1,1}(\Lambda_{\rm R})$ tends to a constant for large $\Lambda_{\rm R}$, while a significant 
drop is observed towards the origin.  In an intermediate range, the opposite signs of 
$c_{0,1}$ and $c_{1,1}$ allow for a compensation of the different effects, implying an 
effectively flat dependence of $\wS$ in the lattice spacing. 
Within the large errors, the mass-dependent discretization effects 
could be compatible with the functional form given in Eq.~\eqref{eq:sigmatildelat}
\cite{Necco:2011vx}. The sign of the pole, however, appears to be opposite 
than predicted in Refs.~\cite{Splittorff:2012gp,Hansen:2011kk}. In this respect 
it must be said that it is not clear that the GSM power-counting scheme
used in Ref.~\cite{Necco:2011vx} applies in the range of parameters of our data.

\bibliographystyle{JHEP}
\bibliography{Literature-cond-nf2.bib}

\end{document}